\documentclass[11pt]{article}

\usepackage{amsmath}
\usepackage{amssymb}
\usepackage{times}
\usepackage{dsfont}

\title{A direct road to Majorana fields}
\author{Andreas Aste\\
Department of Physics, University of Basel, 4056 Basel, Switzerland}
\date{February 3, 2009}

\begin{document}
\maketitle 

\begin{abstract}
\noindent A concise discussion of spin-${1/2}$ field equations with
a special focus on Majorana spinors is presented. The Majorana formalism
which describes massive neutral fermions by the help of two-component or
four-component spinors is of fundamental importance for the understanding
of mathematical aspects of supersymmetric and other extensions of the
Standard Model of particle physics, which may play an increasingly important
role at the beginning of the LHC era. The interplay between the two-component
and the four-component formalism is highlighted in an introductory way.
Majorana particles are predicted both by grand unified theories, in which
these particles are neutrinos, and by supersymmetric theories, in which
they are photinos, gluinos and other states.
\end{abstract}

\section{Introduction}
Starting from Lorentz covariance as one of the key properties of
space-time described in the framework of the special theory of relativity,
we derive the free wave equations for the fundamental spin-${1/2}$
particles in a concise manner and discuss their properties with a
special focus on Majorana spinors \cite{Majorana}, based on the
representation theory of the proper Lorentz group. The two-component
formalism for massive spin-${1/2}$ particles is investigated in detail
and related to the four-component Dirac and Majorana formalism.
This work is intended to give a solid introduction to the basic
mathematical aspects of Majorana fermion fields which constitute
an important aspect of modern neutrino physics.

One important goal of this work is to show that the most general relativistic
field equation for a four-component spinor $\Psi$ can be written in the form
\begin{equation}
i \gamma^\mu \partial_\mu \Psi-\tilde{m}_{_M} \Psi^c - \tilde{m}_{_D} \Psi=0 \label{general2}
\end{equation}
with appropriately chosen Majorana and Dirac mass matrices
$\tilde{m}_{_M}$ and $\tilde{m}_{_D}$. Such a spinor describes a charged
Dirac particle for $\tilde{m}_{_M}=0$ or a pair of Majorana particles.
In the latter case, the theory may contain a physical CP violating phase.

\section{Symmetry}
\subsection{The Lorentz group}
The structure of wave equations is intimately connected
with the Lorentz symmetry of space-time. We therefore
review the most important properties of the Lorentz group
and clarify notational details in this section. Additionally,
basic aspects of the representation theory of the Lorentz group,
which are inevitable for the comprehension of relativistic
wave equations, are derived in detail.

Physical laws are generally assumed to be independent of the observer,
and the underlying symmetry which makes it possible to relate
the points of view taken by different observers is expressed by
the Lorentz group. Gravitational effects will be neglected in the
following by the assumption that space-time is flat.

In two different systems of inertia,
the coordinates of a point in Minkowski space-time measured by two
corresponding observers are related by a proper Lorentz transformation
$\Lambda \! \! \in \! \! \mathcal{L}_+^{\uparrow}=SO^+(1,3)$, which does
not cause time or space reflections, via $x'=\Lambda x$ or
$x'^\mu = \Lambda^\mu_{\, \, \nu} x^\nu$,
with $\Lambda \! \in \! SO^+(1,3)$ defined in further detail below.
For the sake of simplicity, the translational symmetry of space-time
contained in the Poincar\'e group is neglected by the assumption
that the two coordinate systems share their point of origin.

The indefinite Minkoswki inner product $(x,y)$ of two vectors $x$, $y$
is preserved by the Lorentz transformation, i.e.
for $x=(x^0,x^1,x^2,x^3)^{\mbox{\tiny{T}}}=
(x_0,-x_1,-x_2,-x_3)^{\mbox{\tiny{T}}}$ and
$y=(y^0,y^1,y^2,y^3)^{\mbox{\tiny{T}}}=
(y_0,-y_1,-y_2,-y_3)^{\mbox{\tiny{T}}}$ we set
\begin{equation}
(x,y)=
x^{\mbox{\tiny{T}}} gy=x'^{\mbox{\tiny{T}}}gy'=x^{\mbox{\tiny{T}}}
\Lambda^{\mbox{\tiny{T}}} g \Lambda y \quad \forall \, x,y \, ,
\end{equation}
where we make use of the usual rules for matrix multiplication and
consider $x$ as a column vector and $x^{\mbox{\tiny{T}}}$ as a row
vector, or
\begin{equation}
g_{\mu \nu} x^\mu y^\nu=x_\mu y^\mu =x'^\mu y'_\mu =
g_{\mu \nu} x'^\mu y'^\nu=
g_{\mu \nu} \Lambda^\mu_{\, \, \alpha} \Lambda^\nu_{\, \,  \beta}
x^\alpha y^\beta
\end{equation}
with the metric tensor $g$ satisfying
\begin{equation}
g_{\mu \nu}=\bigl( g^{-1} \bigr)^{\mu \nu}=g^{\mu \nu}=
\mbox{diag}(1,-1,-1,-1)=
\left(\begin{array}{rrrr}
 1 &  0 & 0  & 0 \\
 0 & -1 & 0  & 0 \\
 0 &  0 & -1 & 0 \\
 0 &  0 & 0  & -1 
\end{array}\right) \; , 
\end{equation}
\begin{equation}
g^\mu_{\, \, \nu}= g^{\mu \alpha} g_{\alpha \nu}=
\delta^\mu_{\, \, \nu}=
\delta^\mu_\nu \, .
\end{equation}
Therefore, the proper Lorentz group is defined by
\begin{equation}
SO^+(1,3)=\{ \Lambda \! \in \! GL(4,\mathds{R}) \, | \,
\Lambda^{\mbox{\tiny{T}}} g \Lambda=g, \,
\Lambda^0_{\, \, 0} \ge 1, \, \mbox{det} \Lambda = +1\},
\end{equation}
whereas the full Lorentz group $O(1,3)$, consisting of four connected
components, is defined by
\begin{equation}
O(1,3)=\{ \Lambda \! \in \! GL(4,\mathds{R}) \, | \,
\Lambda^{\mbox{\tiny{T}}} g \Lambda=g\}.
\end{equation}
$SO^+(1,3)$ is the identity component in $O(1,3)$, containing
the identity Lorentz transformation expressed by the unit matrix
$\mbox{diag}(1,1,1,1)$, whereas the other three topologically separated
pieces of $O(1,3)$ are the components connected to the time reversal
transformation $\Lambda_{_T}=\mbox{diag}(-1,1,1,1)$, space inversion
$\Lambda_{_P}=\mbox{diag}(1,-1,-1,-1)$ and space-time inversion
$\Lambda_{_{PT}}=\mbox{diag}(-1,-1,-1,-1)$.
Note that from $\Lambda^{\mbox{\tiny{T}}} g \Lambda = g$ follows
${\bigl( \Lambda^{\mbox{\tiny{T}}}} \bigr)^{-1}= g \Lambda g^{-1}$
or
\begin{equation}
{\Bigl[ \bigl(\Lambda^{\mbox{\tiny{T}}} \bigr)^{-1}
\Bigl]}^\mu_{\, \, \, \nu}
=g_{\mu \alpha} \Lambda^\alpha_{\, \, \beta} g^{\beta \nu}
=\Lambda_\mu^{\, \, \nu} \; \; , \quad
\Lambda_\mu^{\, \, \alpha} \Lambda^\nu_{\, \, \alpha}=
\delta^\nu_\mu \, ,
\end{equation}
and from $\Lambda^{-1}= g^{-1} \Lambda^{\mbox{\tiny{T}}} g$ we have correspondingly
$ \bigl(\Lambda^{-1} \bigr)^\mu_{\, \, \nu}=\Lambda_\nu^{\, \, \mu}$,
keeping in mind that the index closer to the matrix symbol
denotes the position of the corresponding matrix element in vertical
direction, if the usual rules for matrix manipulations are presumed.
The $\Lambda$ are not tensors, however, formally
lowering and raising the indices of $\Lambda^\mu_{\, \, \nu}$ has
the effect to generate the matrix (elements) of
$(\Lambda^{\mbox{\tiny{T}}} \bigr)^{-1}$.

\subsection{Two-dimensional irreducible representations of the
Lorentz group}
In order to construct the lowest-dimensional non-trivial
representations of the Lorentz group, we introduce the
quantities
\begin{displaymath}
\bar{\sigma}_\mu=\sigma^\mu=(1,\vec{\sigma}),
\end{displaymath}
\begin{equation}
\bar{\sigma}^\mu=\sigma_\mu=(1,-\vec{\sigma}),
\end{equation}
where the three components of $\vec{\sigma}$ are the Pauli matrices
\begin{equation}
\sigma_1=\left(\begin{array}{rr} 0 & 1 \\
1 & 0 \end{array}\right) \; , \quad
\sigma_2=\left(\begin{array}{rr} 0 & -i \\
i & 0 \end{array}\right) \; , \quad
\sigma_3=\left(\begin{array}{rr} 1 & 0 \\
0 & -1 \end{array}\right) \quad .
\end{equation}
Throughout the paper, we will denote the identity matrix
in two dimensions, alternatively, by the symbols $1$, $I$, or
$\sigma_0$.

From an arbitrary 4-vector $x$ with contravariant
components $x^\mu=(x^0,x^1,x^2,x^3)$ we construct the following
$2 \! \times \! 2$-matrix $X$
\begin{equation}
X=\bar{\sigma}_\mu x^\mu=
\left(\begin{array}{cc} x^0+x^3 & x^1-ix^2 \\
x^1+ix^2 & x^0-x^3 \end{array}\right) \; . \quad
\end{equation}
The map $x \rightarrow X$ is obviously linear and one-to-one,
and $X$ is Hermitian. A further important property of $X$ is
the fact that its determinant is equal to the Minkowski norm
$x^2=(x,x)$
\begin{equation}
\mbox{det} X=(x^0)^2-(x^1)^2-(x^2)^2-(x^3)^2=x_\mu x^\mu.
\end{equation}
The special linear group $SL(2,\mathds{C})$ is defined by
\begin{equation}
SL(2,\mathds{C})=\{S \! \in \! GL(2,\mathds{C}) \, | \,
\mbox{det} S = +1 \}.
\end{equation}
The trick is now to set
\begin{equation}
X'=SXS^+ , \quad S \! \in \! SL(2,\mathds{C}), \label{trick}
\end{equation}
where the $^+$ denotes the Hermitian conjugate matrix.
Then one observes that $X'$ is again Hermitian, and
\begin{equation}
\mbox{det} X' = \mbox{det} (SXS^+)= 
\mbox{det} S \, \mbox{det} X \, \mbox{det} S^+ = \mbox{det} X,
\end{equation}
i.e. $X'$ can again be written as
\begin{equation}
X'=\bar{\sigma}_\mu x'^\mu, \quad \mbox{where} \quad
\quad x'_\mu x'^\mu=x_\mu x^\mu.
\end{equation}
We conclude that $x'$ and $x$ are related by a Lorentz transformation
$x'=\Lambda(S)x$, and since
the group $SL(2,\mathds{C})$ is connected and the map $S \rightarrow
\Lambda(S)$ is continuous, the map $\lambda: S \rightarrow \Lambda(S)$
is a homomorphism of $SL(2,\mathds{C})$ into the proper Lorentz group
$\mathcal{L}_+^{\uparrow}=SO^+(1,3)$.

We now show that the homomorphism $\lambda$ is two-to-one.
This can be easily understood from the observation that $S$ and
$-S$ correspond to the same Lorentz transformation. To see this in more
detail, we calculate the kernel of $\lambda$, i.e. the set of all
$S \! \in \! SL(2,\mathds{C})$ which for any Hermitian matrix $X$
satisfies the equality
\begin{equation}
X=SXS^+ \label{condition}
\end{equation}
By taking, in particular, 
\begin{displaymath}
X=\left(\begin{array}{rr} 1 & 0 \\
0 & 1 \end{array}\right) \; ,
\end{displaymath}
the equality leads to the condition $S=(S^+)^{-1}$, and
eq. (\ref{condition}) reduces to $XS-SX=[X,S]=0$ for any
Hermitian $X$. This implies $S=\alpha I$. From the condition
$\mbox{det} \, S=+1$ follows $S=\pm I$.

Apart from the important group isomorphism
\begin{equation}
\mathcal{L}^\uparrow_+ \cong SL(2,\mathds{C})/\{\pm 1\} ,
\end{equation}
just found, matrices in the special linear group $SL(2,\mathds{C})$ have an
additional interesting feature.
Defining the antisymmetric matrix $\epsilon$ by
\begin{equation}
\epsilon=i \sigma_2=\left(\begin{array}{rr} 0 & 1 \\
-1 & 0 \end{array}\right) \; , \quad \epsilon=-\epsilon^{-1}=
-\epsilon^{\mbox{\tiny{T}}},
\end{equation}
we can also define the symplectic bilinear form $\langle u ,v \rangle
=-\langle v , u \rangle$
for two elements (spinors)
\begin{equation}
u=\left(\begin{array}{cc} u^1 \\ u^2 \end{array}\right) \; , \quad
v=\left(\begin{array}{cc} v^1 \\ v^2 \end{array}\right)
\end{equation}
in the two-dimensional complex vector space
$\mathds{C}^2_{_{\mathds{C}}}$
\begin{equation}
\langle u , v \rangle = u^1 v^2 - u^2 v^1=
u^{\mbox{\tiny{T}}} \epsilon v.
\end{equation}
This symplectic bilinear form is invariant under $SL(2,\mathds{C})$
\begin{equation}
\langle u , v \rangle = u^{\mbox{\tiny{T}}} \epsilon v=
\langle Su , Sv \rangle = u^{\mbox{\tiny{T}}} S^{\mbox{\tiny{T}}}
\epsilon S v.
\end{equation}
Indeed, setting
\begin{equation}
S=\left(\begin{array}{rr} s^1_{\, \, 1} & s^1_{\, \, 2} \\
s^2_{ \, \, 1} & s^2_{ \, \, 2} \end{array}\right) \; , \quad
\mbox{where} \quad
\mbox{det} \, S=s^1_{\, \, 1} s^2_{ \, \, 2}-s^1_{\, \, 2}
s^2_{ \, \, 1},
\end{equation}
a short calculation yields
\begin{displaymath}
S^{\mbox{\tiny{T}}} \epsilon S=
\left(\begin{array}{rr} s^1_{\, \, 1} & s^2_{\, \, 1} \\
s^1_{ \, \, 2} & s^2_{ \, \, 2} \end{array}\right)
\left(\begin{array}{rr} 0 & 1 \\
-1 & 0 \end{array}\right) 
\left(\begin{array}{rr} s^1_{\, \, 1} & s^1_{\, \, 2} \\
s^2_{ \, \, 1} & s^2_{ \, \, 2} \end{array}\right)=
\end{displaymath}
\begin{equation}
\left(\begin{array}{rr} s^1_{\, \, 1} & s^2_{\, \, 1} \\
s^1_{ \, \, 2} & s^2_{ \, \, 2} \end{array}\right)
\left(\begin{array}{cc} s^2_{\, \, 1} & s^2_{\, \, 2} \\
-s^1_{ \, \, 1} & -s^1_{ \, \, 2} \end{array}\right)=
\left(\begin{array}{cc} 0 & \mbox{det} \, S \\
-\mbox{det} \, S & 0 \end{array}\right)=
\left(\begin{array}{rr} 0 & 1 \\
-1 & 0 \end{array}\right) \; ,
\end{equation}
and therefore we have a further group isomorphism
\begin{equation}
\mathcal{L}^\uparrow_+ \cong Sp(2,\mathds{C})/\{ \pm 1 \}, \quad
SL(2,\mathds{C}) \cong Sp(2,\mathds{C}),
\end{equation}
with the complex symplectic group $Sp(2,\mathds{C})$ defined
by
\begin{equation}
Sp(2,\mathds{C})=\{ S \! \in \! GL(2,\mathds{C}) \, | \,
S^{\mbox{\tiny{T}}} \epsilon S=\epsilon \}.
\end{equation}
The group isomorphism $SO(3,\mathds{C}) \cong \mathcal{L}^\uparrow_+$
is mentioned here as a side remark. 

We conclude this section by a crucial observation.
We have learned above how to construct an action of the Lorentz
group on the two-dimensional complex vector space
$\mathds{C}^2_{_\mathds{C}}$, which obviously respects by definition
the Lorentz group structure in the sense of representations
\begin{equation}
\Lambda(S_1) \Lambda(S_2) = \Lambda(S_1 S_2).
\end{equation}
Up to a sign, this relation can be inverted and one may write in
a slightly sloppy style
$S(\Lambda_1 \Lambda_2)=\pm S(\Lambda_1) S(\Lambda_2)$ to
express the fact that we have found a double-valued representation
of the Lorentz group.
One might wonder whether this representation is equivalent to
the representation given by the
complex conjugate special linear matrices $S^*(\Lambda)$.
This is not the case, since it is impossible to relate
all $S \! \in \! SL(2,\mathds{C})$ by a basis transform
such that
\begin{equation}
S^*=B S B^{-1} \quad \forall \, S.
\end{equation}
However, considering only the subgroup
\begin{equation}
SU(2)=\{ U \! \in \! GL(2,\mathds{C}) \, | \, U^+=U^{-1}, \,
\mbox{det} \, U=1 \} \! \subset \! SL(2,\mathds{C}),
\end{equation}
the situation is different.
A special unitary matrix $U \! \in \! SU$ can be written in the
form
\begin{equation}
U=\left(\begin{array}{cc} a & b \\
-b^* & a^* \end{array}\right) \; , \quad
\mbox{det} \, U=aa^*+bb^*=1,
\end{equation}
and a short calculation yields
\begin{equation}
U^*=\epsilon U \epsilon^{-1},
\end{equation}
with the $\epsilon$ defined above.
The two-dimensional two-valued representation of the group of
rotations $SO(3)$ by special unitary transformations, which can
be obtained directly via the trick eq. (\ref{trick}), is equivalent
to its complex conjugate representation.

\section{Field equations}
\subsection{The scalar field}
\subsubsection{Wave equation}
Before we reach our goal to construct wave equations for
spinor fields, we shortly review the simplest case of
a free (non-interacting) scalar field $\varphi(x)$ satisfying
the Klein-Gordon equation ($\hbar=c=1$)
\begin{equation}
\{ \Box + m^2 \} \varphi(x)=
\{ \partial_\mu \partial^\mu + m^2 \} \varphi(x)=0, \label{kg1}
\end{equation}
which is given in the primed coordinate system by
\begin{equation}
{ \varphi'}(x')=\varphi(x)=\varphi(\Lambda^{-1} x')=\varphi'(\Lambda x).
\end{equation}
The field ${ \varphi'}(x')$ indeed satisfies the
Klein-Gordon equation also in the primed coordinate system
\begin{equation}
\{ \Box' + m^2 \} \varphi'(x')=
\{ \partial'_\mu \partial'^\mu + m^2 \} { \varphi'}(x')=0,
\label{kg2}
\end{equation}
since the differential operators $\partial_\mu$, $\partial^\mu$
transform according to
\begin{equation}
\partial_\nu=\frac{\partial}{\partial^\nu}=
\frac{\partial x'^\mu}{\partial x^\nu}  \frac{\partial}{\partial x'^\mu}=
\Lambda^\mu_{\, \, \nu} \frac{\partial}{\partial x'^\mu}=
\Lambda^\mu_{\, \, \nu} \partial'_\mu \, , \quad \mbox{or}
\end{equation}
\begin{equation}
\partial'_\alpha=\delta_\alpha^\mu \partial'_\mu=\Lambda_\alpha^{\, \, \nu}
\Lambda^\mu_{\, \, \nu} \partial'_\mu\ = \Lambda_\alpha^{\, \, \nu}
\partial_\nu,
\end{equation}
and is a analogous manner one derives $\partial'^\mu=\partial/\partial
x'_\mu=\Lambda^\mu_{\, \, \nu} \partial^\nu$.
Hence we have
\begin{equation}
\partial'_\mu \partial'^\mu { \varphi'}(x')=
\Lambda_\mu^{\, \, \alpha}  \Lambda^\mu_{\, \, \beta}
\partial_\alpha  \partial^\beta { \varphi'}(\Lambda x)=
g^\alpha_\beta \partial_\alpha \partial^\beta \varphi(x)=
\partial_\alpha \partial^\alpha \varphi(x),
\end{equation}
i.e. eqns. (\ref{kg1},\ref{kg2}) are equivalent.

Note that a non-trivial Lorentz-invariant first order
differential operator $\sim \alpha_\mu \partial^\mu, \,
\alpha_\mu \! \in \! \mathds{C}$,
acting on a one-component field does not
exist, such that the Klein-Gordon equation is necessarily of
second order.

\subsubsection{Charge conjugation}
Positive and negative energy plane-wave solutions of the
Klein-Gordon equation are given by
\begin{equation}
\varphi(x)=e^{\pm ikx}, \quad kx=k_\mu x^\mu,
\quad k^0=\sqrt{\vec{k}^2+m^2},
\end{equation}
and the general solution of the Klein-Gordon is a superposition
of plane waves which can be expressed by a Fourier integral \`a la
\begin{equation}
\varphi(x)=\int \frac{d^3 k}{2 k^0 (2 \pi)^3} [a(\vec{k}) e^{-ikx}+b^*(\vec{k})
e^{+ikx}],
\end{equation}
when the field is complex, or by
\begin{equation}
\varphi(x)=\int \frac{d^3 k}{2 k^0 (2 \pi)^3} [a(\vec{k}) e^{-ikx}+a^*(\vec{k})
e^{+ikx}],
\end{equation}
when the field is real. Since the positive energy solutions
$e^{-ikx}=e^{i \vec{k} \vec{x}-i k^0 x_0}$ can be associated with
particles and the negative energy solutions
$e^{+ikx}=e^{-i \vec{k} \vec{x}+i k^0 x_0}$
with antiparticles, it is natural to define charge conjugation
for scalar C-number fields by
\begin{equation}
\varphi^c(x)=\eta_c \varphi^*(x),
\end{equation}
since this transformation does interchange the role of particle
and antiparticle solutions and leaves the momentum (and the spin,
if particles with spin are considered) of particles
unchanged. We set the phase factor $\eta_c$, $|\eta_c|=1$, which is
typically chosen to be $\pm1$ for a neutral field, equal to one
in the following.

By quantization, the Fourier coefficients
$a,a^*$ and $b,b^*$ become particle creation or annihilation operators
$a,a^+$ and $b,b^+$,
complex conjugation corresponds then to Hermitian conjugation $^+$.
Within this formalism, the charge conjugation operator $U_c$
then acts on $\varphi$ as unitary operator such that
$\varphi^c(x)=U_c \varphi(x) U_c^{-1} = \varphi^+(x)$.
This can easily be accomplished if we define
\begin{equation}
U_c a^+(\vec{k}) U_c^{-1}=b^+(\vec{k}),
\end{equation}
\begin{equation}
U_c b^+(\vec{k}) U_c^{-1}=a^+(\vec{k}).
\end{equation}
Thus, the charge conjugation operator acts on a
one-particle $a^+(\vec{k}) |0 \rangle$
state with momentum $\vec{k}$
\begin{equation}
U_c a^+(\vec{k}) |0 \rangle = U_c a^+(\vec{k}) U_c^{-1} U_c |0 \rangle
=U_c a^+(\vec{k}) U_c^{-1} |0 \rangle = b^+ (\vec{k}) |0 \rangle,
\end{equation}
where we presupposed that the vacuum $|0 \rangle=U_c |0 \rangle$
is invariant under charge conjugation.

We observe that charge conjugation can be discussed to some extent on a
C-number field level, and we will proceed that way when we come
to the fermion fields. However, one should always keep in mind
that after second quantization, subtleties may arise.
Note that quantum mechanical states live in a complex
Hilbert space and can be gauged by complex phase factors.
From a naive point of view, we may consider a real field as a
"wave function" or quantum mechanical state, but then we are not
allowed to multiply the field by an imaginary number, since the
field is then no longer a real entity.
One may also remark that the widespread labeling of antiparticle
solutions as "negative energy solutions" is misleading to some extent and
should be replaced rather by "negative frequency solutions" in the
literature, since the energy of antiparticles is positive.

From two (equal mass) degenerate real fields one may construct
a complex field
\begin{equation}
\tilde{\varphi}=\frac{1}{\sqrt{2}}(\varphi_1+i \varphi_2) ,
\end{equation}
and the corresponding creation and annihilation operators
$a_1$, $a_1^+$, $a_2$, and $a_2^+$ can be combined, e.g., to
\begin{equation}
a=\frac{1}{\sqrt{2}}(a_1+i a_2), \quad
b^+=\frac{1}{\sqrt{2}}(a_1^+ + ia_2^+) \neq a^+.
\end{equation}
Vice versa, the two real fields can be recovered from the
complex field. One should not consider the real particles
as bound states of the charged states, since no interaction
is present. The process described above is of a purely
formal nature.

Dirac particles can be constructed in an analogous way
from two neutral Majorana fields, as we shall see later.

\subsection{Weyl equations}
We now turn to wave equations for two-component wave wave
functions, and derive the Weyl equations for massless fermions.
A linear first-order differential operator acting on a spinor
is given by
\begin{equation}
{\hat{\sigma}}={\bar{\sigma}}_\mu \partial^\mu=
\left(\begin{array}{cc} \partial^0+\partial^3 & \partial^1-i \partial^2 \\
\partial^1+i \partial^2 & \partial^0-\partial^3
\end{array}\right) \; , \quad
\end{equation}
If we apply a Lorentz transformation to this object, then
$\hat{\sigma}$ obviously transforms according to
\begin{equation}
\hat{\sigma}'={\bar{\sigma}}_\mu \partial'^\mu=
\Lambda^\mu_{\, \, \nu} {\bar{\sigma}}_\mu \partial^\nu=
S \hat{\sigma} S^+ ,
\end{equation}
according to the trick given by eq. (\ref{trick}).
If we apply $\hat{\sigma}$ to a (column) spinor
\begin{equation}
\psi(x)=\left(\begin{array}{cc} \psi_{1} (x) \\
\psi_{2} (x) \end{array}\right) \; ,
\end{equation}
we obtain the wave equation
\begin{equation}
\hat{\sigma} \psi=(\sigma_0 \partial_0 - \vec{\sigma} \vec{\nabla})
\psi=0. \label{Weyl1}
\end{equation}
This is one of the famous Weyl equations \cite{Weyl}. However, we have to check
whether the differential equation also holds in all systems of inertia.
This is true if the spinor $\psi$ transform according to
\begin{equation}
\psi(x) \rightarrow \psi'(x')=
\epsilon S^* \epsilon^{-1} \psi(\Lambda^{-1}x'), \label{trafo_law}
\end{equation}
i.e. according to a representation which is equivalent to the
complex conjugate representation constructed in the previous section.
Since all special linear matrices fulfill $S^{\mbox{\tiny{T}}}
\epsilon S=\epsilon$, we also have $(S^{\mbox{\tiny{T}}})^{-1}=
\epsilon S \epsilon^{-1}$ and $(S^+)^{-1}=\epsilon S^* \epsilon^{-1}$,
and consequently
\begin{equation}
\hat{\sigma}' \psi' =
S \hat{\sigma} S^+ (\epsilon S^* \epsilon^{-1} \psi)=
S \hat{\sigma} S^+ (S^{-1})^+ \psi=S \hat{\sigma} \psi=0,
\label{Weyl1a}
\end{equation}
since $\hat{\sigma} \psi=0$.
A similar equation can be obtained if we use the differential
operator
\begin{equation}
\check{\sigma}=\epsilon \hat{\sigma} \epsilon^{-1}=
(\sigma_0+{\vec{\sigma}}^{\mbox{\tiny{T}}} \vec{\nabla})=
\sigma_\mu^* \partial^\mu
\end{equation}
and a (row) spinor
\begin{equation}
\tilde{\psi}(x)^{\mbox{\tiny{T}}}
=\bigl( \tilde{\psi}_1 (x) \, \, \tilde{\psi}_2 (x) \bigr).
\end{equation}
A wave equation is given by
\begin{equation}
\tilde{\psi}(x)^{\mbox{\tiny{T}}} \check{\sigma}=0, \label{Weyl2}
\end{equation}
equivalent to the transposed equation
\begin{equation}
\check{\sigma}^{\mbox{\tiny{T}}} \tilde{\psi}(x)=
(\sigma_0 \partial_0 + \vec{\sigma} \vec{\nabla}) \tilde{\psi}(x)=0,
\end{equation}
which is just the Weyl equation with differing sign compared to
eq. (\ref{Weyl1}).
The spinor transforms according to
\begin{displaymath}
\tilde{\psi}(x) \rightarrow S \tilde{\psi}(\Lambda^{-1}x'),
\end{displaymath}
\begin{equation}
\tilde{\psi}(x)^{\mbox{\tiny{T}}} \rightarrow
\tilde{\psi}(\Lambda^{-1}x')^{\mbox{\tiny{T}}} S^{\mbox{\tiny{T}}}, \label{trafo_law_right}
\end{equation}
such that
($\check{\sigma} \rightarrow \epsilon S \hat{\sigma} S^+ \epsilon^{-1}$,
$S^{\mbox{\tiny{T}}} \epsilon S=\epsilon$)
\begin{equation}
\tilde{\psi}^{\mbox{\tiny{T}}} S^{\mbox{\tiny{T}}} \epsilon
S \hat{\sigma} S^+ \epsilon^{-1}=
\tilde{\psi}^{\mbox{\tiny{T}}} \epsilon \hat{\sigma} 
(\epsilon^{-1} \epsilon) S^+ \epsilon^{-1}=
\tilde{\psi}^{\mbox{\tiny{T}}} \check{\sigma} \cdot
\epsilon S^+ \epsilon^{-1}=
\tilde{\psi}^{\mbox{\tiny{T}}} \check{\sigma} \cdot
(S^*)^{-1}=0.
\end{equation}

Originally, the Weyl equations were considered unphysical
since they are not invariant under space reflection. Considering the
parity transformation
\begin{equation}
\Lambda_{_P}: (x^0,\vec{x}) \rightarrow (x^0,-\vec{x}), \quad
\psi'(x)=\psi(x^0,-\vec{x}),
\end{equation}
the transformed version of eq. (\ref{Weyl1}) which reads
$(\sigma_0 \partial_0+\vec{\sigma} \vec{\nabla}) \psi(x)=0$
would be equivalent if there existed a linear isomorphic transformation
$S_{_P}$ of the spinor with
\begin{equation}
S_{_P} (\sigma_0 \partial_0+\vec{\sigma} \vec{\nabla}) \psi'(x)=
(\sigma_0 \partial_0-\vec{\sigma} \vec{\nabla}) S_{_P} \psi'(x)
\end{equation}
for all solutions $\psi(x)$ of the Weyl equation, which implies
$\sigma_k S_{_P}=-S_{_P} \sigma_k$, for $k=1,2,3$. But
this is only possible for $S_{_P}=0$

A further apparent defect is the absence of a mass term
$\sim m \psi$ or $\sim m \tilde{\psi}$ in the Weyl equations,
since these terms behave differently under a Lorentz transformation than
the related differential operator part $\hat{\sigma} \psi$ or
$\check{\sigma} \tilde{\psi}$.
However, this statement is a bit overhasty, as we shall see in the
following section. 

\subsection{Two-component Majorana fields}
\subsubsection{Majorana wave equations}
\label{two_component}
We reach now our goal and construct the wave equations for
free Majorana fields.
We add a mass term to the Weyl equation given by eq. (\ref{Weyl1})
or eq. (\ref{Weyl1a}) by observing that the complex conjugate
spinor $\psi^*(x)$ transforms according to
\begin{equation}
\psi^*(x) \rightarrow \epsilon S \epsilon^{-1} \psi^*(\Lambda^{-1}x'),
\end{equation}
and therefore the {\emph{Majorana mass term}}
$m \epsilon^{-1} \psi^*(x)$ transforms like the differential operator
part of the Weyl equation
\begin{equation}
m \epsilon^{-1} \psi^*(x) \rightarrow
m \epsilon^{-1} \epsilon S \epsilon^{-1} \psi^*(\Lambda^{-1}x')=
S(m \epsilon^{-1} \psi^*(\Lambda^{-1}x')),
\end{equation}
such that the equation
\begin{equation}
\hat{\sigma} \psi(x)=\pm i m \epsilon^{-1} \psi^*(x)
\end{equation}
or
\begin{equation}
(\sigma_0 \partial_0 - \vec{\sigma} \vec{\nabla})
\psi(x) \mp i m \epsilon^{-1} \psi^*(x)=0
\end{equation}
is indeed Lorentz invariant. The imaginary unit in front
of the mass term has been chosen in order to respect
manifestly the {\emph{CP invariance}} of the theory for the moment, however,
more general considerations will be presented below.
Note that the sign of the mass term in the present framework is a matter of taste,
since from a solution $\psi(x)$ of the wave equation one obtains
directly $i \psi(x)$ as a solution of the analogous equation with
flipped mass.

The same trick works for the alternative Weyl equation eq. (\ref{Weyl2}).
There we have
\begin{equation}
\tilde{\psi}^*(x) \rightarrow S^* \tilde{\psi} (\Lambda^{-1} x'),
\end{equation}
and ($\epsilon S^+ \epsilon^{-1}=(S^*)^{-1}$)
\begin{equation}
m \tilde{\psi}^{*{\mbox{\tiny{T}}}} \epsilon^{-1} \rightarrow
m \tilde{\psi}^{*{\mbox{\tiny{T}}}} S^{*{\mbox{\tiny{T}}}} \epsilon^{-1}=
m \tilde{\psi}^{*{\mbox{\tiny{T}}}} \epsilon^{-1} \epsilon
S^+ \epsilon^{-1}=
m \tilde{\psi}^{*{\mbox{\tiny{T}}}} \epsilon^{-1} \cdot (S^*)^{-1},
\end{equation}
leading to the alternative equation
\begin{equation}
(\sigma_0 \partial_0 + \vec{\sigma} \vec{\nabla})
\tilde{\psi}(x) \pm i m \epsilon^{-1} \tilde{\psi}^*(x)=0.
\end{equation}

For both equations, we choose a definite mass sign convention
and obtain the left-chiral and right-chiral
{\emph{two-component Majorana equations}}
\begin{equation}
i (\sigma_0 \partial_0-\vec{\sigma} \vec{\nabla}) \psi(x) +
m \epsilon^{-1} \psi(x)^*=0, \label{Majo1}
\end{equation}
\begin{equation}
i (\sigma_0 \partial_0+\vec{\sigma} \vec{\nabla}) \tilde{\psi}(x) -
m \epsilon^{-1} \tilde{\psi}(x)^*=0, \label{Majo2}
\end{equation}
which can also be written
\begin{equation}
i {\bar{\sigma}}^\mu \partial_\mu \psi(x) - m(i \sigma_2) \psi^*(x)=0,
\end{equation}
\begin{equation}
i \sigma^\mu \partial_\mu {\tilde{\psi}}(x) + m(i \sigma_2) {\tilde{\psi}}^*(x)=0.
\end{equation}
Eq. (\ref{Majo1}) may be expressed in matrix form ($x,y,z
\sim 1,2,3$)
\begin{equation}
i \left(\begin{array}{cc} \partial_0-\partial_z  & -\partial_x+i
\partial_y+im K \\
-\partial_x -i \partial_y-im K & \partial_0+\partial_z
\end{array}\right)
\left(\begin{array}{c} \psi_{1} (x) \\
\psi_{2} (x)
\end{array}\right) =0,
\end{equation}
where the operator $K$ denotes the complex conjugation.
Multiplying this equation by
\begin{equation}
-i \left(\begin{array}{cc} \partial_0+\partial_z  & \partial_x-i
\partial_y+im K \\
\partial_x +i \partial_y-im K & \partial_0-\partial_z
\end{array}\right)
\end{equation}
from the left, one obtains
\begin{equation}
\left(\begin{array}{cc}\Box+m^2  & 0 \\
0 & \Box+m^2
\end{array}\right)
\left(\begin{array}{c} \psi_{1} (x) \\
\psi_{2} (x)
\end{array}\right) =0,
\end{equation}
i.e. the components of the Majorana spinor $\psi$ and in an
analogous way the components of $\tilde{\psi}$ obey the Klein-Gordon
equation, and it is a straightforward task to construct plane
wave solutions of the Majorana equation.

A first important remark should be made concerning the
existence of two Majorana equations. As we have seen above,
the existence of two non-equivalent equations eq. (\ref{Majo1}) and
eq. (\ref{Majo2}) is related
to the fact that a two-dimensional complex spinor can transform
in two different ways under Lorentz transformations.
This fundamental property of a spinor, whether it transforms according to
the representation $S(\Lambda)$ or $S(\Lambda)^*$, is called
{\emph{chirality}}. This phenomenon also exists at a simpler level
for the group $U(1)=\{z \! \in \! \mathds{C} \, | \, zz^*=1 \}$.
There are two true representations given by
$\exp(i \alpha)=z \mapsto z$ and $\exp(i \alpha) \mapsto \exp(-i \alpha)$.
Spinor fields are not physical observables, however, their chirality
is important when one aims at the construction of observables using
spinorial entities.

A further critical remarks should be made about the charge conjugation
properties of the Majorana fields constructed above.
Starting from
\begin{equation}
(\sigma_0 \partial_0-\vec{\sigma} \vec{\nabla}) \psi(x)
-im \epsilon^{-1} \psi^*(x)=0, \label{majostart}
\end{equation}
inserting $\epsilon^{-1} \epsilon=\sigma_0$ in front of the spinor
and multiplying the equation from the left by $\epsilon$, we obtain
\begin{equation}
\epsilon (\sigma_0 \partial_0-\vec{\sigma} \vec{\nabla})
\epsilon^{-1} \epsilon \psi(x)
-im \epsilon^{-1} \epsilon \psi^*(x)=
(\sigma_0 \partial_0+\vec{\sigma}^* \vec{\nabla}) \epsilon \psi(x)
-im \epsilon^{-1} \epsilon \psi^*(x)=0
\end{equation}
or, equivalently, the complex conjugate equation
\begin{equation}
(\sigma_0 \partial_0+\vec{\sigma} \vec{\nabla}) \epsilon \psi^*(x)
+im \epsilon^{-1} \epsilon \psi(x)=0. \label{Majo_conj}
\end{equation}
The new field $\epsilon \Psi^*$ does not fulfill the original
Majorana equation, however, charge conjugation acting on wave
functions is by definition an antilinear operator which does not
alter the direction of the momentum and the spin of a particle.
Since there also exists no linear transformation of the
field which would remedy this situation, we have to accept that
charge conjugation defined here by $\psi \rightarrow \epsilon \psi^*$
transforms the field out of its equivalence class.

Although the signs of the spatial derivatives and the mass term
differ from the original equation, we may
perform a parity transformation including a multiplication of the spinor
$\epsilon \psi^*(x^0,-\vec{x})$ by the {\emph{CP eigenphase}} $i$.
Of course, a CP eigenphase with opposite sign would also suffice.
Then we have
\begin{equation}
(\sigma_0 \partial_0-\vec{\sigma} \vec{\nabla}) (i \epsilon
\psi^*(x^0,-\vec{x}))
-im \epsilon^{-1} (i \epsilon \psi^*(x^0,-\vec{x}))^*=0.
\end{equation}
Therefore, the {\emph{CP conjugate}} spinor $CP[\psi(x^0,\vec{x})]=\pm i \epsilon
\psi^*(x^0,-\vec{x})$ is again a solution of the original
Majorana equation, and a CP transformed particle
can be converted by a Lorentz transformation into its original
state. This arguments no longer holds in the massless Weyl case.

Additionally, one should keep in mind that CP conjugation is not
an exact symmetry in nature. This implies that the story becomes
even more involved when interactions start to play a role.
Therefore, the actual {\emph{definition of a Majorana particle}}
is given by the demand that a Majorana particle be an "eigenstate"
of the CPT transformation, where T denotes the time reversal operator
introduced below. Note, however, that the notion "CPT eigenstate",
which is widely found in the literature, should not be misunderstood
in the narrow sense that a particle state remains invariant under CPT.
E.g., CPT changes the spin of a particle. However, a CPT transformed
particle state is again a physical state of the same type of particle.
Furthermore, CPT expresses a fundamental symmetry of every local
quantum field theory.
We shall see below that the definition of a Majorana particle
formulated above is contained in a natural manner in the
two-component Majorana formalism.

We finally mention that a scalar complex field has {\emph{two}} two charge
degrees of freedom, whereas a Majorana field has {\emph{two}}
polarization degrees of freedom. This observation is a starting
point for supersymmetric theories, where fermions and bosons
get closely related.

\subsubsection{CPT}
Based on eq. (\ref{majostart}), we find that the field
$\psi(-x^0,\vec{x})$ fulfills the equation
\begin{equation}
(-\sigma_0 \partial_0-\vec{\sigma} \vec{\nabla})
\psi(-x^0,\vec{x})
-im \epsilon^{-1} \psi^*(-x^0,\vec{x})=0.
\end{equation}
Using analogous tricks as above, this equation can be converted into
\begin{equation}
(\sigma_0 \partial_0-\vec{\sigma} \vec{\nabla}) ( \epsilon
\psi^*(-x^0,\vec{x}))
-im \epsilon^{-1} ( \epsilon \psi^*(-x^0,\vec{x}))^*=0.
\end{equation}
Defining the time reversal transformation according to
$T[\psi(x^0,\vec{x})]=\epsilon \psi^*(-x^0,\vec{x})$, we obtain
$CPT[\psi(x^0,\vec{x})]= \pm i\epsilon(\epsilon \psi^*(-x^0,-\vec{x}))^*
=\pm i \epsilon^2 \psi(-x^0,-\vec{x})=
\mp i \psi(-x^0,-\vec{x})$,
and the CPT transformed Majorana field again fulfills the original
wave equation eq. (\ref{Majo1}) or eq. (\ref{majostart}).

There is a crucial point in this discussion above.
While the charge conjugation C and CP are related to symmetries which
are violated by some interactions, in local relativistic quantum field theory all
particle interactions respect the CPT symmetry \cite{Luders}.
This means that to every particle process in nature
there is an associated CPT conjugate process with properties that
can be inferred exactly from the original system. Thus, we emphasize again
that the basic property which makes a fermion a Majorana particle
is the fact that a particle state with definite four-momentum and spin
can be transformed by a CPT transformation and a subsequent (space-time) 
Poincar\'e transformation into itself.
In a world, where neither C, P, T, CP, CT, nor PT are conserved,
this is the only way of expressing the fact that a particle is its own
antiparticle. A physical Majorana neutrino, subject to maximally C-violating
weak interactions, cannot be an eigenstate of C. It may be an "approximate
eigenstate" of CP.

\subsubsection{Plane wave solutions}
\label{Solutions}
In the previous section, we tacitly chose phase conventions and the
mass term in the Majorana equations such that the resulting fields
displayed most simple transformations properties under CP conjugation.
We consider now the most general mass term for one type of
spin-$1/2$ Majorana particles and calculate explicit
solutions of the non-interacting left-chiral Majorana equation
\begin{equation}
i (\sigma_0 \partial_0-\vec{\sigma} \vec{\nabla}) \psi(x) =
\eta m \epsilon \psi(x)^*,  \label{Majo_phase}
\end{equation}
where $\eta=e^{i \delta}$ is a phase $|\eta|=1$.
In the non-interacting case, the complex mass term $\eta m$ has no physical
impact and could be rendered real in eq. (\ref{Majo_phase})
by globally gauging the wave function $\psi(x) \rightarrow
\psi'(x)=\psi(x) e^{-i \delta/2}$, such that
\begin{displaymath}
i (\sigma_0 \partial_0-\vec{\sigma} \vec{\nabla}) \psi(x)
-\eta m \epsilon \psi(x)^*=
i (\sigma_0 \partial_0-\vec{\sigma} \vec{\nabla}) e^{i \frac{\delta}{2}} \psi'(x)
-e^{i \delta} m \epsilon e^{-i \frac{\delta}{2}} \psi'(x)^*
\end{displaymath}
\begin{equation}
=e^{i \frac{\delta}{2}} [i (\sigma_0 \partial_0-\vec{\sigma} \vec{\nabla}) \psi'(x)
-m \epsilon \psi'(x)^*]=0.
\end{equation}
When interactions are involved, the discussion of Majorana phases becomes
important \cite{Mohapatra}.

Eq. (\ref{Majo_phase}) follows from the Lagrangian ($\psi^+=\psi^{\mbox{\tiny{T}}*}$)
\begin{equation}
\mathcal{L}= \psi^+ i \bar{\sigma}^\mu \partial_\mu \psi
-\eta \frac{m}{2} \psi^+ (i \sigma_2) \psi^*
+\eta^* \frac{m}{2} \psi^{\mbox{\tiny{T}}} (i \sigma_2) \psi,
\end{equation}
which contains so-called Majorana mass terms. Note that the mass terms above
equal zero, if the components of the Majorana field are assumed to be
ordinary numbers. However, in the present fermionic case, we adopt the rule
the classical fermionic fields {\emph{anticommute}}.

For particles at rest, eq. (\ref{Majo_phase}) can be written as
\begin{equation}
i \dot{\psi}_1=\eta m \psi_2^* , \quad i \dot{\psi}_2= -\eta m \psi_1^*.
\end{equation}
Differentiating the equation on the left and using the equation on
the right $\dot{\psi}_2^*=-i \eta^* m \psi_1$,
one obtains
\begin{equation}
\ddot{\psi}_1=-i \eta m \dot{\psi}_2^* = -|\eta|^2 m^2 \psi_1
=-m^2 \psi_1.
\end{equation}
$\psi_1$ is therefore a linear combination of $e^{-imx^0}$ and
$e^{+imx^0}$, and solutions describing a particle with spin
parallel or antiparallel to the 3- or z-direction are given
by ($\psi_2=-\frac{i}{m} \eta \dot{\psi}_1^*$)
\begin{equation}
\psi_{+\frac{1}{2}}=\left(\begin{array}{c} 1 \\ 0  \end{array}\right)
e^{-imx^0}+
\eta \left(\begin{array}{c} 0 \\ -1  \end{array}\right) e^{+imx^0},
\end{equation}
\begin{equation}
\psi_{-\frac{1}{2}}=\left(\begin{array}{c} 0 \\ 1  \end{array}\right)
e^{-imx^0}+
\eta \left(\begin{array}{c} 1 \\ 0  \end{array}\right) e^{+imx^0}.
\end{equation}
Plane wave solutions always contain a positive and negative energy (or, rather,
{\emph{frequency}}) part, which combine to describe an uncharged particle.
As a hand-waving argument, one may argue that the "negative energy"
part in the solutions displayed above corresponds to a particle
hole and must therefore be equipped with an opposing spinor.

The phase $\eta$ has no physical meaning as long as 
weak interactions or mixings of different particles are absent,
field operators related to the negative energy part $\sim e^{+imx^0}$
of the wave functions above can then
be redefined such that the phase disappears.
Otherwise, this phase will appear as one of the so-called
Majorana phases, potentially in conjunction with other phases
related to charged Dirac fermions.

Solutions for moving particles can be generated by boosting
the solutions given above according to the transformation law
eq. (\ref{trafo_law}). The corresponding matrices are given here
without derivation
\begin{equation}
S=\sqrt{\frac{E+m}{2m}} \Biggl( 1+ \frac{\vec{\sigma} \vec{k}}{E+m} \Biggr),
\end{equation}
\begin{equation}
\epsilon S^* \epsilon^{-1}=
\sqrt{\frac{E+m}{2m}} \Biggl( 1- \frac{\vec{\sigma} \vec{k}}{E+m} \Biggr).
\end{equation}
A Majorana particle with momentum $\vec{k}$ and spin
in z-direction is described, e.g., by
\begin{equation}
\psi_{+\frac{1}{2}}(\vec{k},x)=\epsilon S^* \epsilon^{-1}
\left(\begin{array}{c} 1 \\ 0  \end{array}\right)
e^{-ikx}+ \eta \epsilon S^* \epsilon^{-1}
\left(\begin{array}{c} 0 \\ -1  \end{array}\right) e^{+ikx},
\quad kx=k_\mu x^\mu.
\end{equation}

\subsubsection{Helicity}
In the following, we focus on an alternative representation
of two-component Majorana spinors.
Given a momentum $\vec{k}$ in a direction specified by the polar
coordinates $\theta$ and $\phi$
\begin{equation}
\vec{k}=|\vec{k}| (\sin \theta \cos \phi, \sin \theta \sin \phi,
\cos \theta),
\end{equation}
the Pauli spin states \cite{Pauli}
\begin{equation}
h_+ =
\left(\begin{array}{c}
+\cos (\theta/2) e^{-i \phi/2} \\ +\sin (\theta/2) e^{+i \phi/2}
\end{array}\right), \quad
h_- =
\left(\begin{array}{c}
-\sin (\theta/2) e^{-i \phi/2} \\ +\cos (\theta/2) e^{+i \phi/2}
\end{array}\right)
\end{equation}
are helicity eigenstates
\begin{equation}
\vec{\sigma} \vec{k} h_{\pm} = \pm |\vec{k}| | h_\pm,
\end{equation}
where
\begin{equation}
\vec{\sigma} \vec{k}=|\vec{k}| 
\left(\begin{array}{cc}
\cos \theta & \sin \theta e^{-\phi} \\
\sin \theta e^{+i \phi} &  -\cos \theta
\end{array}\right).
\end{equation}
The Pauli spinors are obviously related by ($\epsilon= i \sigma_2$)
\begin{equation}
h_+^c=\epsilon h_+^*=-h_- , \quad h_-^c=\epsilon h_-^*=+ h_+ .
\end{equation}
In order to solve the Majorana equation eq. (\ref{Majo1})
\begin{equation}
i(\sigma_0 \partial_0 - \vec{\sigma} \vec{\nabla}) \psi (x)=m \epsilon \psi^* (x),
\label{Majo_helicity}
\end{equation}
we make the Ansatz
\begin{equation}
\psi(x)=[\alpha_+ h_+ e^{-ikx} + {\tilde{\alpha}}_+ h_- e^{+ikx}] +
[\alpha_- h_- e^{-ikx}+{\tilde{\alpha}}_- h_+ e^{+ikx}]. \label{ansatz_helicity}
\end{equation}
Note that a negative frequency solution
$\sim e^{+ikx}$ with a negative helicity spinor $h_-$ corresponds to a particle with positive
helicity, which clarifies the meaning of the indices of the coefficients above.
Inserting the Ansatz eq. (\ref{ansatz_helicity}) in eq. (\ref{Majo_helicity}) leads
to the relations (with $E=k^0=(\vec{k}^2 +m^2)^{1/2}$, and $K=|\vec{k}|$ in this section)
\begin{equation}
+\alpha_+ (E+K)= +{\tilde{\alpha}}_+^* m, \quad 
{\tilde{\alpha}}_+ (-E+K)=-\alpha_+^* m,
\end{equation}
\begin{equation}
+\alpha_- (E-K) = - {\tilde{\alpha}}_-^* m, \quad
-{\tilde{\alpha}}_- (E+K) = + \alpha_-^* m,
\end{equation}
where $\vec{\sigma} (-i \vec{\nabla}) e^{\mp ikx}= \pm \vec{\sigma} \vec{k}
e^{\mp ikx}$ was also used.
Up to normalization factors, these conditions imply
\begin{equation}
\alpha_+ = +\sqrt{E-K} e^{+i \delta}, \quad {\tilde{\alpha}}_+=+\sqrt{E+K} e^{-i \delta},
\end{equation}
\begin{equation}
\alpha_- = +\sqrt{E+K} e^{+i \delta'}, \quad {\tilde{\alpha}}_-= -\sqrt{E-K} e^{-i \delta'},
\end{equation}
with $\delta, \delta' \! \in \! \mathds{R}$.

Hence, a general Fourier representation of a free Majorana field can be written as
\begin{displaymath}
\psi(x)=\int \frac{d^3 k}{2 k^0 (2 \pi)^3} 
\Bigl\{ [+\sqrt{E-K} \alpha_+(\vec{k}) h_+ e^{-ikx} + \sqrt{E+K} \alpha_+^* (\vec{k})
h_- e^{+ikx}]+
\end{displaymath}
\begin{equation}
\quad \quad \quad \quad \quad \quad \quad \quad \quad
[+ \sqrt{E+K} \alpha_-(\vec{k}) h_- e^{-ikx} - \sqrt{E-K} \alpha_-^* (\vec{k})
h_+ e^{+ikx}] \Bigr\}. \label{expansion}
\end{equation}
After second quantization, $\alpha_+$, $\alpha_-$, $\alpha_+^*$, and $\alpha_-^*$
become the creation and annihilation operators for the Majorana particle
in the $\pm-$helicity states.

Note that the square root terms above reappear in the matrix elements
describing, e.g., the beta decay of the neutron. In this process,
the right-handed antineutrino is produced predominantly along with
the electron, the amplitude being of the order $\sqrt{E_\nu+K_\nu}
\simeq \sqrt{2 E_\nu}$. The left-handed neutrino has a much smaller
amplitude $\sqrt{E_\nu-K_\nu} \simeq m_\nu/ \sqrt{2 E_\nu}$.
In the massless case, only positive-helicity particles would be produced. Similarly,
in the high-energy limit only the terms containing $\alpha_+^*$
and $\alpha_-$ in eq. (\ref{expansion}) survive, which correspond
to positive-helicity (right-handed) negative energy particles and
left-handed positive energy particles, respectively.
These two possibilities correspond to two helicity states of
a neutral Majorana neutrino, or to the right-handed antineutrino and
left-handed neutrino, respectively. Both cases can only be distinguished if
interactions are present.

\section{The Dirac equation}
\subsection{Representations}
The Dirac equation is the well-known  partial differential equation
of first order, which has been used very successfully for the
description, e.g., of electrons and positrons since its formulation
by Paul Dirac in 1928 \cite{Dirac}.
The equation for the 4-component spinor $\Psi$ describing
interaction-free Dirac particles reads ($\hbar=c=1$)
\begin{equation}
\{ i \gamma^\mu \partial_\mu - m \} \Psi(x),
\end{equation}
where $x=(x^0,x^1,x^2,x^3)=(x_0,-x_1,-x_2,-x_3)$ are the space-time
coordinates and the $\gamma^\mu$ the famous Dirac matrices
$\gamma_{Dirac}=:\gamma^\mu$,
satisfying the anti-commutation relations
\begin{equation}
\{ \gamma^\mu , \gamma^\nu \} = 2 g^{\mu \nu}. \label{commutator}
\end{equation}

There is a standard choice for the Dirac matrices in the literature,
given by ($i=1,2,3$)
\begin{equation}
\gamma^0=\left(\begin{array}{rr} 1 & 0 \\
0 & -1 \end{array}\right) \; , \quad
\gamma^k=\left(\begin{array}{cc} 0 & \sigma_k \\
-\sigma_k & 0 \end{array}\right) \; , \quad
\gamma_5=\gamma^5=i \gamma^0 \gamma^1 \gamma^2 \gamma^3 =
\left(\begin{array}{rr} 0 & 1 \\
1 & 0 \end{array}\right) \quad ,
\end{equation}
where the $1$ stands for the $2 \times 2$ identity matrix.

An important result from the theory of Clifford algebras states
that every set of matrices ${\tilde{\gamma}}^\mu$ 
satisfying the anti-commutation relations eq. (\ref{commutator}) is
equivalent to the standard choice defined above in the sense
that
\begin{equation}
{\tilde{\gamma}}^\mu = U \gamma^\mu U^{-1}
\end{equation}
for some suitable invertible matrix $U$.
This is a nice feature of the gamma matrices, since it makes
sure that two physics communities living in different
solar systems can easily compare their calculations
by a simple transformation. In this sense, the Dirac equation
is unique.

For the so-called chiral (or {\emph{Weyl}}) representation, one has
\begin{equation}
\gamma^0_{ch(iral)}=\gamma^0_{ch}=\left(\begin{array}{rr} 0 & 1 \\
1 & 0 \end{array}\right) \; , \quad
\gamma^k_{ch}=\left(\begin{array}{cc} 0 & -\sigma_k \\
\sigma_k & 0 \end{array}\right) \; , \quad
\gamma^5_{ch}=\gamma_5^{ch}=\left(\begin{array}{rr} 1 & 0 \\
0 & -1 \end{array}\right) \quad ,
\end{equation}
with
\begin{equation}
\gamma^\mu_{chiral}=U \gamma^\mu_{Dirac} U^{-1} \; , \quad
U=\frac{1}{\sqrt{2}} \left(\begin{array}{rr} 1 & 1 \\
1 & -1 \end{array}\right) \;\;\;, \quad U^{-1}=U^+,
\end{equation}
where the $^+$ denotes the Hermitian conjugate.
The chiral Dirac matrices can also be written as
\begin{equation}
\gamma^\mu_{ch}=\left(\begin{array}{rr} 0 & {\bar{\sigma}}^\mu \\
\sigma^\mu & 0 \end{array}\right) \quad .
\end{equation}

The charge conjugation of a Dirac spinor is given in the chiral
representation by $\psi^c=\eta_c \tilde{C} \bar{\psi}^{\mbox{\tiny{T}}}$,
$\bar{\psi}^{\mbox{\tiny{T}}}=\gamma^{0 \mbox{\tiny{T}}}_{ch} \psi^*$,
with $\eta_c$ an arbitrary unobservable phase, generally taken a being
equal to unity and the matrix $\tilde{C}$ is given by
\begin{equation}
\tilde{C}=\left(\begin{array}{rr} i \sigma_2 & 0 \\
0 & -i \sigma_2 \end{array}\right) \; , 
\end{equation}
or
\begin{equation}
C[\psi]=\psi^c = \left(\begin{array}{rr} 0 & \epsilon \\
-\epsilon & 0 \end{array}\right) \psi^*. \label{CCchiral}
\end{equation}
The definition eq. (\ref{CCchiral}) of charge conjugation
is equally valid in the standard representation.

The Majorana representation of Dirac matrices is given by
\begin{displaymath}
i\gamma^0_M=\left(\begin{array}{rrrr}
0 & 0 & 0 & 1 \\
0 & 0 & -1 & 0 \\
0 & 1 & 0 & 0 \\
-1 & 0 & 0 & 0 
\end{array}\right) \;, \quad
i\gamma^1_M=\left(\begin{array}{rrrr}
-1 & 0 & 0 & 0 \\
0 & 1 & 0 & 0 \\
0 & 0 & -1 & 0 \\
0 & 0 & 0 & 1 
\end{array}\right) \;,
\end{displaymath}

\begin{equation}
i\gamma^2_M=\left(\begin{array}{rrrr}
0 & 0 & 0 & -1 \\
0 & 0 & 1 & 0 \\
0 & 1 & 0 & 0 \\
-1 & 0 & 0 & 0 
\end{array}\right) \;, \quad
i\gamma^3_M=\left(\begin{array}{rrrr}
0 & 1 & 0 & 0 \\
1 & 0 & 0 & 0 \\
0 & 0 & 0 & 1 \\
0 & 0 & 1 & 0 
\end{array}\right) ,
\end{equation}
where
\begin{equation}
\gamma^\mu_M=\gamma^\mu_{Majorana}=V \gamma^\mu_{Dirac} V^+, \quad
V=V^{-1}=V^+=\frac{1}{\sqrt{2}}
\left(\begin{array}{cc}
1 & \sigma_2 \\
\sigma_2 & -1 
\end{array}\right) . \label{Vtransform}
\end{equation}
Note that the Majorana-Dirac matrices are purely imaginary, such that
the Dirac equation contains only real coefficients. Therefore, it is possible
to enforce that the solutions of the equation are purely real.
Such solutions correspond to a Majorana field, since the
charge conjugation operator in the Majorana representation is given
simply by the complex conjugation of all spinor components.
This basic definition of charge conjugation, which is motivated by
the fact that both a given solution and its complex conjugate
fulfill the same Dirac equation with real coefficients, can still
be modified by a phase.

The definition of charge conjugation by eq. (\ref{CCchiral})
conforms to the charge conjugation defined in a rather heuristic
manner in Sect. \ref{two_component} for two-component spinors in
an obvious way.
This correspondence can also be achieved for the parity transformation.
Defining the parity transformed Dirac spinor in an arbitrary representation
by
\begin{equation}
\Psi^p(x'^0,\vec{x}')=\gamma^0 \Psi(x^0,\vec{x}), \quad x'^0=x^0, \quad \vec{x}'=-\vec{x},
\end{equation}
one finds that the new spinor satisfies the Dirac equation
as well, since
\begin{equation}
\gamma^\mu \partial'_\mu \Psi^p(x')=
(\gamma^0 \partial'_0 -\gamma^k \partial'_k) \gamma^0 \Psi(x^0,\vec{x})=
(\gamma^0 \partial_0 +\gamma^k \partial_k) \gamma^0 \Psi(x^0,\vec{x})=
\gamma^\mu \partial_\mu \Psi (x),
\end{equation}
where we used that $\gamma^k \gamma^0=-\gamma^0 \gamma^k$ for $k=1,2,3$.

For a detailed discussion of the time reversal operator we refer to the literature.
Apart from an admissible phase factor, it is given in the Dirac representation
for a four-component spinor by
\begin{equation}
\Psi(x^0,\vec{x}) \rightarrow T[\Psi(x^0,\vec{x})]= i \gamma^3 \gamma^5 \Psi(-x^0,\vec{x})^*
=  - \gamma^2 \gamma^5 \gamma^0 \Psi(-x^0,\vec{x})^*.
\label{time_reversal}
\end{equation}
Note that both the replacement $x^0 \rightarrow -x^0$ and the complex conjugation
in eq. (\ref{time_reversal}) flip the positive and negative frequency part
of a plane wave solution of the Dirac equation. As a consequence, in the case of
a C-symmetric theory based on, e.g., the Dirac equation describing non-interacting
charged fermions like electrons and positrons, the charge of the particles is
invariant under a time reversal transformation. However, spin and momentum change sign.
The reader should also keep in mind that the definitions used in the present
work for the charge, parity and time reversal transformations change their formal
structure when they are considered in the framework of (second quantized)
quantum field theories. It should also be pointed out that the definitions presented
in this section are constructed in a manner that is compatible with the
corresponding definitions originally given for two-component spinors.

\subsection{Chiral decomposition of the Dirac and Majorana equation}
Decomposing the Dirac spinor in the chiral representation
into two two-component spinors
\begin{equation}
\Psi=\left(\begin{array}{cc}
\Psi_R \\ \Psi_L
\end{array}\right) \quad ,
\end{equation}
the Dirac equation becomes
\begin{equation}
i \left(\begin{array}{cc}
0 & \sigma_0 \partial_0 -\vec{\sigma} \vec{\nabla} \\
\sigma_0 \partial_0 +\vec{\sigma} \vec{\nabla} & 0  
\end{array}\right)
\left(\begin{array}{cc}
\Psi_R \\ \Psi_L 
\end{array}\right)-
\left(\begin{array}{cc}
m & 0 \\ 0 & m   \end{array}\right)
\left(\begin{array}{cc}
\Psi_R \\ \Psi_L 
\end{array}\right)=0,
\end{equation}
or
\begin{equation}
(\sigma_0 \partial_0 -\vec{\sigma} \vec{\nabla}) \Psi_L+im \Psi_R=0,
\end{equation}
\begin{equation}
(\sigma_0 \partial_0 +\vec{\sigma} \vec{\nabla}) \Psi_R+im \Psi_L=0.
\end{equation}
The Dirac equation describes two fields
with different chirality, which are coupled by mass
terms. In the Standard Model, only $\Psi_L$-fields take part in the
electroweak interaction. Without mass term in the Dirac equation,
$\Psi_L$ would describe a Dirac neutrino field in the
(obsolete version of the) Standard Model, where
neutrinos are assumed to be massless.

Defining the four-component Majorana spinor by
\begin{equation}
\Psi_M=
\left(\begin{array}{c}
{\tilde{\psi}} \\ \psi
\end{array}\right) \quad ,
\end{equation}
the two-component Majorana equations eq. (\ref{Majo1}) and eq. (\ref{Majo2})
can be cast into the four-component {\emph{Majorana equation}}
\begin{equation}
i \gamma^\mu \partial_\mu \Psi_M - m \Psi_M^c=0. \label{Majo4_double}
\end{equation}
This way we absorb both the left- and the right-chiral two-component Majorana fields
in one four-spinor, describing indeed 4 degrees of freedom.
This way one may construct a new type of Majorana field which describes indeed
charge neutral particles, a strategy, which will be used below.
It is, however, always possible to extract left- and right-chiral
fields from the Majorana spinor above by using the corresponding projection
operators.
Actually, the mass term in the Majorana equation can be generalized
such that it describes neutral particles with two different masses
$m_1$ and $m_2$, as discussed in Sect. \ref{masses}.

\subsection{CP violating phase}
The observation that there are different types of spin-$1/2$
particles mights raise the question whether the Dirac equation is
really unique up to different choices of the Dirac
matrices. In fact, it is not. The modified Dirac equation
\begin{equation}
i \gamma^\mu \partial_\mu \psi(x)-m e^{i \Theta' \gamma^5} \psi(x)=0
\label{chiral_dirac}
\end{equation}
is not in conflict with relativistic covariance. However,
a non-trivial so-called chiral phase $\Theta'$ may lead to CP violating effects
when interactions are present, due to its properties under parity
transformation or charge conjugation which involves complex conjugation.
Note that the phase discussed here is in close analogy to the phase which appeared
already above in eq. (\ref{Majo_phase}).

It is straightforward to verify that the mass of the mass eigenstates
following from eq. (\ref{chiral_dirac}) is given by the parameter $m$,
also in close analogy to the discussion presented in Sect. \ref{Solutions}.
The chiral mass term $\tilde{m}=m e^{i \Theta' \gamma^5}$ should not be confused
with the effective complex mass of a decaying particle $M+i \Gamma/2$.
The $\gamma^5$-part of the chiral mass term
appearing in the Hamiltonian corresponding to eq. (\ref{chiral_dirac})
is indeed Hermitian and respects unitary time evolution of the Dirac wave function.
Defining a chirally transformed wave function
\begin{equation}
\psi'(x)=e^{+i \frac{\Theta'}{2} \gamma^5} \psi(x), \label{chiral_gauge}
\end{equation}
one first observes that
\begin{equation}
e^{+i \frac{\Theta'}{2} \gamma^5} \gamma^\mu = \gamma^\mu e^{-i \frac{\Theta'}{2} \gamma^5},
\end{equation}
since $\gamma^5$ anticommutes with the Dirac matrices: $\gamma^5 \gamma^\mu=-\gamma^\mu \gamma^5$ for
$\mu=0,...,3$. Consequently, $\psi'(x)$ fulfills the ordinary Dirac equation with a real mass, since
\begin{displaymath}
i \gamma^\mu \partial_\mu \psi'(x)-m \psi'(x)=
i \gamma^\mu \partial_\mu e^{+i \frac{\Theta'}{2} \gamma^5} \psi(x)-
m e^{-i \frac{\Theta'}{2} \gamma^5}  e^{+i \Theta' \gamma^5} \psi(x)
\end{displaymath}
\begin{equation}
=e^{-i \frac{\Theta'}{2} \gamma^5} \bigl[
i \gamma^\mu \partial_\mu \psi(x)-
m e^{+i \Theta' \gamma^5} \psi(x) \bigr]=0,
\end{equation}
and $\psi(x)$ fulfills eq. (\ref{chiral_dirac}).
From this observation, one should not conclude that the chiral phase
can be simply rotated away in theories with interactions like the Standard Model.
There, the chiral phase of the quarks is linked with the so-called $\Theta$-Term,
which might be present in the gluonic part of the QCD Lagrangian, via the so-called triangle
anomaly \cite{Stein,Adler,Bell}. Furthermore, chiral rotations are not a symmetry
of the full theory. The combined $\Theta$- and $\Theta'$ terms
are related to a potential electric dipole moment of the neutron in a highly non-trivial way
\cite{Mitra}.

\subsection{Four-component Majorana fields}
In the following, we will rename the field $\psi$ obeying
eq. (\ref{Majo1}) by $\chi_{_L}$ and $\tilde{\psi}$ obeying
eq. (\ref{Majo2}) by $\chi_{_R}$, in order to stress their
transformation properties under $SL(2,\mathds{C})$.
It is a bit unfortunate that the symbols $L$ and $R$, which
denote the chirality of the fields, might suggest a connection
to the helicity (or handedness) of particles, which is not
a conserved quantity in the massive case.

We construct additionally a four-component spinor
\begin{equation}
\nu_{_L}=\left(\begin{array}{c} 0  \\
\chi_{_L}  \end{array}\right) \; , 
\end{equation}
such that we can write eq. (\ref{Majo1}) in the {\emph{chiral}}
representation as follows
\begin{equation}
i \left(\begin{array}{cc} 0 & \partial_0-\vec{\sigma} \vec{\nabla} \\
\partial_0+\vec{\sigma} \vec{\nabla} & 0 \end{array}\right)
\left(\begin{array}{c} 0  \\
\chi_{_L}  \end{array}\right)-
\left(\begin{array}{cc} m & 0 \\
0 & m \end{array}\right)
\left(\begin{array}{c} \epsilon \chi^*_{_L} \\
0  \end{array}\right) =0,
\end{equation}
or
\begin{equation}
i \gamma^\mu_{ch} \partial_\mu \nu_{_L}-m \nu_{_L}^c=
i \gamma^\mu_{ch} \partial_\mu \nu_{_L}-m (\nu_{_L})^c
=0, \quad
\frac{1}{2}(1-\gamma^5_{ch}) \nu_{_L}=\nu_{_L}, \label{wave_left}
\end{equation}
where the symbol $c$ denotes charge conjugation as it is defined
in the chiral representation. We proceed one step further and
define a {\emph{neutral}} four-component field
\begin{equation}
\nu^{_M}_{_1}=\nu_{_L}+\nu_{_L}^c=\left(\begin{array}{c}  \epsilon \chi^*_{_L} \\
\chi_{_L}  \end{array}\right),
\quad \nu^{_M}_{_1}={\nu^{_M}_{_1}}^c.
\end{equation}
We have seen in the section above that $\epsilon \psi^*$ fulfills
eq. (\ref{Majo_conj}), hence we have
\begin{equation}
i \gamma_{ch}^\mu \partial_\mu \nu^{_M}_{_1} -m \nu^{_M}_{_1}  =0, \quad \nu^{_M}_{_1}
={\nu^{_M}_{_1}}^c. \label{4Majorana}
\end{equation}
This four-component Majorana field is often used in
supersymmetric theories, and the same construction
naturally works for $\nu_2=\chi_{_R}+\chi_{_R}^c$. A formalism using
$\nu_1$'s only must be equivalent to one using $\nu_2$'s only.
The fields $\nu_1$ and $\nu_2$ are charge conjugation invariant
by construction, and it is common practice to denote this type of fields
as the actual Majorana fields. The free chiral Majorana fields
can readily be recovered by from the charge self-conjugate
Majorana fields by projections $\nu_{R,L}=\frac{1}{2} (1 \pm \gamma^5_{ch})
\nu_{1,2}$. One notational advantage of the four-component formalism relies on
the fact that the description of interactions of Majorana neutrinos
with Dirac particles naturally involves expressions
based on Dirac four-component spinors.

Eq. (\ref{wave_left}) is equivalent to
\begin{equation}
i \gamma^\mu_{ch} \partial_\mu \nu_{_L}^c-m \nu_{_L} =0,
\end{equation}
hence we have
\begin{equation}
i \gamma^\mu_{ch} \partial_\mu (\nu_{_L}-\nu_{_L}^c)-m (\nu_{_L}^c-\nu_{_L}) =0,
\end{equation}
or
\begin{equation}
i \gamma^\mu_{ch} \partial_\mu [i(\nu_{_L}-\nu_{_L}^c)]-m [i(\nu_{_L}-\nu_{_L}^c)]^c=
i \gamma^\mu_{ch} \partial_\mu [i(\nu_{_L}-\nu_{_L}^c)]-m [i(\nu_{_L}-\nu_{_L}^c)]=0.
\end{equation}
The same construction works for $\nu_{_R}$,
and we may also construct the following combinations of Majorana
(quantum) fields
\begin{equation}
\psi_1=\frac{1}{\sqrt{2}}(\nu_{_L}+\nu_{_L}^c+\nu_{_R}+\nu_{_R}^c), \quad
\psi_1=\psi_1^c,
\end{equation}
\begin{equation}
\psi_2=\frac{i}{\sqrt{2}}(\nu_{_L}-\nu_{_L}^c+\nu_{_R}-\nu_{_R}^c), \quad
\psi_2=\psi_2^c.
\end{equation}
We do not discuss normalization factors in the present qualitative discussion on
a first-quantized level.
Combining the two physically distinct fields given above leads to a Dirac field $\psi_{_D}$
\begin{equation}
\psi_{_D}=\frac{1}{\sqrt{2}} (\psi_1-i\psi_2) \sim \nu_{_L}+\nu_{_R},
\end{equation}
which is no longer neutral, since $\psi_{_D}^c \neq \psi_{_D}^c$.

Note that each Majorana field $\nu^{_M}_{_1}$ or $\nu^{_M}_{_2}$ from eq. (\ref{4Majorana})
is completely determined by two complex quantities
\begin{equation}
\nu^{_M}_{_1}=
\left(\begin{array}{c}  \epsilon \chi^*_{_L} \\
\chi_{_L}  \end{array}\right)=
\left(\begin{array}{c}  \chi^*_{_{L_2}} \\ -\chi^*_{_{L_1}} \\
\chi_{_{L_1}} \\ \chi_{_{L_2}} \end{array}\right),
\qquad
\nu^{_M}_{_2}=
\left(\begin{array}{c}  \chi_{_R} \\ -\epsilon \chi^*_{_R} 
\end{array}\right)=
\left(\begin{array}{c}  \chi^{_1}_{_R} \\ \chi^{_2}_{_R} \\
-\chi^{_2*}_{_R} \\ \chi^{_1*}_{_R} \end{array}\right),
\end{equation}
whereas a Dirac spinor has 4 complex entries, such that the four-component
Majorana fields constructed above have only two physical degrees of freedom
instead of four. We will see below that we can represent Majorana spinors
in a purely real form, where the two complex spinor components
$\chi_{L_1}$ and $\chi_{L_2}$ correspond to four real numbers.

The fact that it is possible to construct charge conjugation eigenstates
in the four-component formalism should be considered as a convenient mathematical
trick. We emphasize again that CPT is the fundamental symmetry of
local relativistic quantum field theories, whereas C and CP are not.
To give a rather intuitive picture of the meaning of the CPT symmetry,
we point out that in an even-dimensional Euclidean space, a point reflection
of the space can also be obtained from a continuous rotation of the space
itself. This is impossible for the space-time reflection PT in Minkowski
space. Therefore, we have no reason to assume that for every physical process,
there exists a space-time reflected process. However, the requirement
of locality {\emph{and}} relativistic invariance in quantum field theory is strong
enough to ensure that an additional transformation C can always be found such
that to every physical process, a C-PT mirrored process exists.
The C-conjugate state of a physical particle does not necessarily exist,
it may be sterile in the sense of non-interacting.
The "true" antiparticle is related to its partner via the CPT transformation,
and both the particle and the antiparticle have the same mass.
All these observations necessitate a generalization of the considerations
made so far. 

\subsection{A comment on gauge invariance}
A very important property of a charged Dirac field describing, e.g.,
electrons and positrons, which is coupled to a vector potential $A_\mu$
describing the electromagnetic field, is the gauge invariance
of the Dirac equation including the coupling to the gauge field
\begin{equation}
i \gamma^\mu (\partial_\mu+ie A_\mu) \Psi - m \Psi=0, \label{DiracA}
\end{equation}
where $e$ is the negative charge of the electron.
The gauge transformed fields
\begin{equation}
\Psi'(x)=e^{-ie \eta (x)}\Psi(x),
\end{equation}
\begin{equation}
A'_\mu(x)=A_\mu(x)+\partial_\mu \eta(x),
\end{equation}
still fulfill the equation of motion eq. (\ref{DiracA}),
and also the Maxwell equations with the electric current
source term remains invariant.

A glimpse at eqns. (\ref{Majo1}) and (\ref{Majo2}) clearly
shows that a corresponding gauge transformation is impossible
for Majorana fields, since the complex conjugate spinor in
the mass term acquires the wrong phase such that the gauge
transformed spinor no longer fulfills the Majorana equation.
Therefore, a Majorana particle is truly neutral.
This situation changes abruptly, when the Majorana mass term
vanishes.

It is common practice to define charge conjugation for the
vector potential by
\begin{equation}
A_\mu^c=-A_\mu \label{cc_potential}
\end{equation}
since then electric and magnetic field also change sign as one expects
when positive and negative charges in a physical system are
exchanged. To complicate the situation, one could additionally invoke a gauge
transformation of the vector potential which does not change the
physically observable electric and magnetic fields.
A full discussion of charge conjugation (and other
discrete transformations as space reflection and time reversal)
is indeed much more involved, since the quantization of gauge
potentials is not trivial. An introduction on some aspects
of symmetry transformations in quantum field theory is given by
\cite{Kemmer}.

Accepting the simple convention eq. (\ref{cc_potential}), the
charge conjugate version of the interacting Dirac equation
\begin{equation}
i \gamma^\mu (\partial_\mu+ie A_\mu) \Psi - m \Psi=0
\end{equation}
is 
\begin{equation}
[i \gamma^\mu (\partial_\mu+ie A_\mu) \Psi - m \Psi]^c=
i \gamma^\mu (\partial_\mu-ie A_\mu) \Psi^c - m \Psi^c=0.
\end{equation}
It is therefore clearly forbidden that $\Psi=\Psi^c$ for a
Dirac field.

\section{Real forms of Majorana fields}
\subsection{Generators of the Lorentz group and the unimodular
group}
A pure Lorentz boost in $x^1$-direction with velocity
$v=\beta c$ is expressed by the
matrix
\begin{equation}
\Lambda^\mu_{\, \, \nu} =\left(\begin{array}{cccc}
\gamma & -\gamma \beta & 0 & 0 \\
-\gamma \beta & \gamma & 0 & 0 \\
0 & 0 & 1 & 0 \\
0 & 0 & 0 & 1 
\end{array}\right) \;, \quad \gamma=\frac{1}{\sqrt{1-\beta^2}}, \quad
\gamma^2-\gamma^2 \beta^2=1.
\end{equation}
For $\beta \ll 1$, we can write to first order in $\beta$
\begin{equation}
\Lambda^\mu_{\, \, \nu} =\left(\begin{array}{cccc}
1 & 0 & 0 & 0 \\
0 & 1 & 0 & 0 \\
0 & 0 & 1 & 0 \\
0 & 0 & 0 & 1 
\end{array}\right)+
\beta
\left(\begin{array}{cccc}
0 & -1 & 0 & 0 \\
-1 & 0 & 0 & 0 \\
0 & 0 & 0 & 0 \\
0 & 0 & 0 & 0 
\end{array}\right) =1+\beta K_1 , 
\end{equation}
where $K_1$ is a generator for boosts in $x^1$-direction,
and from the theory of continuous groups it is well-known that
the original Lorentz boost can be recovered by
exponentiating the generator
\begin{equation}
\exp (\xi K_1)=\left(\begin{array}{cccc}
+\cosh \xi & -\sinh \xi & 0 & 0 \\
-\sinh \xi & +\cosh \xi & 0 & 0 \\
0 & 0 & 1 & 0 \\
0 & 0 & 0 & 1 
\end{array}\right) \; , \quad \cosh(\xi)=\gamma,
\end{equation}
with an appropriately chosen boost parameter $\xi$.
For boosts in $x^2$- and $x^3$-direction, one has
\begin{equation}
K_2=\left(\begin{array}{cccc}
0 & 0 & -1 & 0 \\
0 & 0 & 0 & 0 \\
-1 & 0 & 0 & 0 \\
0 & 0 & 0 & 0 
\end{array}\right)\; , \quad
K_3=\left(\begin{array}{cccc}
0 & 0 & 0 & -1 \\
0 & 0 & 0 & 0 \\
0 & 0 & 0 & 0 \\
-1 & 0 & 0 & 0 
\end{array}\right)\; , \quad
\end{equation}
and the generators for rotations around the $x^1$-, $x^2$-, and
$x^3$-axis are given by
\begin{equation}
S_1=\left(\begin{array}{cccc}
0 & 0 & 0 & 0 \\
0 & 0 & 0 & 0 \\
0 & 0 & 0 & -1 \\
0 & 0 & 1 & 0 
\end{array}\right)\; , \quad
S_2=\left(\begin{array}{cccc}
0 & 0 & 0 & 0 \\
0 & 0 & 0 & 1 \\
0 & 0 & 0 & 0 \\
0 & -1 & 0 & 0 
\end{array}\right)\; , \quad
S_3=\left(\begin{array}{cccc}
0 & 0 & 0 & 0 \\
0 & 0 & -1 & 0 \\
0 & 1 & 0 & 0 \\
0 & 0 & 0 & 0 
\end{array}\right)\; . \quad
\end{equation}
The six generators of the proper Lorentz group given above
satisfy the commutation relations
\begin{equation}
[S_l,S_m]=+\varepsilon_{lmn} S_n, \quad
[K_l,K_m]=-\varepsilon_{lmn} S_n, \quad
[K_l,S_m]=+\varepsilon_{lmn} K_n, 
\end{equation}
with $\varepsilon_{123}=1$, $\varepsilon_{lmn}=-\varepsilon_{mln}
=-\varepsilon_{lnm}$ the totally antisymmetric tensor
in three dimensions. Note that the generators above get often
multiplied with the imaginary unit $i$ in the physics literature
in order to get Hermitian matrices.

The Lie algebra $so(1,3)$ of the proper Lorentz group
is isomorphic to the Lie algebra $sl(2,\mathds{C})$, and it is not
difficult to find a basis of generators in $sl(2,\mathds{C})$
which span the real six-dimensional vector space
such that an infinitesimal Lorentz transformation
\begin{equation}
1+\nu_l S_l+\beta_l K_l
\end{equation}
corresponds one-to-one to an infinitesimal spinor transformation
\begin{equation}
1+\nu_l \tilde{S}_l+\beta_l \tilde{K}_l
\end{equation}
in accordance with eq. (\ref{trick}).
The generators in $sl(2,\mathds{C})$ are given by
\begin{equation}
\tilde{S}_k=-i \sigma_k, \quad \tilde{K}_k=\sigma_k,
\end{equation}
and since the Pauli matrices fulfill the anti-commutation relation
\begin{equation}
[\sigma_l,\sigma_m]=i \varepsilon_{lmn} \sigma_n,
\end{equation}
one readily verifies that
\begin{equation}
[\tilde{S}_l,\tilde{S}_m]=+\varepsilon_{lmn} \tilde{S}_n, \quad
[\tilde{K}_l,\tilde{K}_m]=-\varepsilon_{lmn} \tilde{S}_n, \quad
[\tilde{K}_l,\tilde{S}_m]=+\varepsilon_{lmn} \tilde{K}_n. \label{slcomm}
\end{equation}

\subsection{The real spinor representation of the Lorentz group in four
dimensions}
The $sl(2,\mathds{C})$-generators presented above are matrices
with complex elements, and it is clearly impossible so
satisfy the algebraic relations eq. (\ref{slcomm}) with matrices
containing only real elements.
Interestingly, there exists a real representation of the
group $SL(2,\mathds{C})$ in four dimensions.
Instead of giving a detailed derivation, we directly list
a set of real generators with all demanded properties.
We denote the boost generators by $K^r_1$, $K^r_2$, $K^r_3$,
which are given by
\begin{equation}
\frac{1}{2} \left(\begin{array}{cccc}
0 & 0 & 0 & -1 \\
0 & 0 & -1 & 0 \\
0 & -1 & 0 & 0 \\
-1 & 0 & 0 & 0 
\end{array}\right)\, , \, \,
\frac{1}{2} \left(\begin{array}{cccc}
1 & 0 & 0 & 0 \\
0 & 1 & 0 & 0 \\
0 & 0 & -1 & 0 \\
0 & 0 & 0 & -1 
\end{array}\right)\, , \, \,
\frac{1}{2} \left(\begin{array}{cccc}
0 & 0 & -1 & 0 \\
0 & 0 & 0 & 1 \\
-1 & 0 & 0 & 0 \\
0 & 1 & 0 & 0 
\end{array}\right)\, , \, \,
\end{equation}
respectively, and the rotations are generated by
$S^r_1$, $S^r_2$, $S^r_3$, given by
\begin{equation}
\frac{1}{2} \left(\begin{array}{cccc}
0 & 0 & 1 & 0 \\
0 & 0 & 0 & -1 \\
-1 & 0 & 0 & 0 \\
0 & 1 & 0 & 0 
\end{array}\right)\, , \, \,
\frac{1}{2} \left(\begin{array}{cccc}
0 & -1 & 0 & 0 \\
1 & 0 & 0 & 0 \\
0 & 0 & 0 & -1 \\
0 & 0 & 1 & 0 
\end{array}\right)\, , \, \,
\frac{1}{2} \left(\begin{array}{cccc}
0 & 0 & 0 & -1 \\
0 & 0 & -1 & 0 \\
0 & 1 & 0 & 0 \\
1 & 0 & 0 & 0 
\end{array}\right)\, ,
\end{equation}
respectively.
An explicit calculation illustrates that the matrices above indeed
generate a double-valued spinor representation of the
proper Lorentz group, since
\begin{equation}
\exp(2 \pi \nu_l S^r_l)=-1 \quad \mbox{for} \, \,
\nu_1^2+\nu_2^2+\nu_3^2=1,
\end{equation}
i.e. one full physical rotation of a spinor around an arbitrary axis $\vec{\nu}$
changes the sign of the spinor.

Although the {\emph{irreducible}} real double-valued representation
of the proper Lorentz group constructed above
acts on the four-dimensional real space like the proper Lorentz group
itself, one should keep in mind that it clearly has completely different geometrical
properties.

\subsection{Real Majorana spinors}
We finally construct real Majorana spinors in a concrete manner, starting
from the solutions of the Dirac equation in the standard representation.

Using for the moment the standard representation of Dirac matrices,
plane wave solutions describing electrons (particles) with their spin in $z-$direction
and momentum $\vec{k}$ are given, up to normalization factors, by
\begin{equation}
u_s(\vec{k},x)=
\left(\begin{array}{c}  \xi_s \\
\frac{\vec{\sigma} \vec{k}}{E+m} \xi_s  \end{array}\right) e^{-ikx}, \quad
s=\pm \frac{1}{2}, \quad \xi_{+\frac{1}{2}}=\left(\begin{array}{c}  1 \\
0 \end{array}\right), \quad
\xi_{-\frac{1}{2}}=\left(\begin{array}{c}  0 \\
1 \end{array}\right),
\end{equation}
whereas positrons (antiparticles) with their spin parallel or antiparallel to the
$z-$axis and momentum $\vec{k}$ are described by the
charge conjugate plane wave spinors ($\epsilon \vec{\sigma}^*=-\vec{\sigma} \epsilon$)
\begin{equation}
v_s(\vec{k},x)=i \gamma^2 u_s(\vec{k},x)^*=u_s(\vec{k},x)^c=
\left(\begin{array}{c}
\frac{\vec{\sigma} \vec{k}}{E+m} \epsilon \xi_s  \\
\epsilon \xi_s
\end{array}\right) e^{+ikx},
\end{equation}
where charge conjugation is defined by eq. (\ref{CCchiral})
and $E=k^0=(\vec{k}^2+m^2)^{1/2}$ is the (positive) energy of the particles.

If we combine particle and antiparticle solutions corresponding to
the same spin and momentum according to
\begin{equation}
w_s(\vec{k},x)=\frac{1}{\sqrt{2}} (u_s(\vec{k},x)+i v_s(\vec{k},x)), \label{Dirac_neutral}
\end{equation}
we obtain {\emph{neutral}} charge conjugation eigenstates with eigenvalue $-i$, since
\begin{equation}
w_s(\vec{k},x)^c=\frac{1}{\sqrt{2}} (u_s(\vec{k},x)^c-i v_s(\vec{k},x)^c)=
-i w_s(\vec{k},x).
\end{equation}
Why the $v-$spinor has been equipped with an imaginary prefactor will become clear below.
From $w_s$, we construct new spinors $m_s$ which are solutions of
the Dirac equation in the Majorana representation.
This is achieved by a linear transformation
\begin{equation}
m_s(\vec{k},x)=V w_s(\vec{k},x)
\end{equation}
using
the matrix $V$ from eq. (\ref{Vtransform}), which connects the standard and
Majorana representation.
An explicit calculation yields ($x \sim x^\mu$, $(x^1,x^2,x^3) \sim (x,y,z)$)
\begin{equation}
m_{+\frac{1}{2}}(\vec{k},x)=
\left(\begin{array}{c}  +\cos(kx)-\frac{1}{E+m} (k_x \sin(kx)-k_y \cos(kx)) \\
+\frac{k_z}{E+m} \sin(kx) \\
-\frac{k_z}{E+m} \cos(kx) \\
+\sin(kx) - \frac{1}{E+m} (k_x \cos(kx)+k_y \sin(kx))
\end{array}\right) \quad ,
\end{equation}
\begin{equation}
m_{-\frac{1}{2}}(\vec{k},x)=
\left(\begin{array}{c}  +\frac{k_z}{E+m} \sin(kx)  \\
+\cos(kx)+\frac{1}{E+m} (k_x \sin(kx)+k_y \cos(kx)) \\
-\sin(kx) - \frac{1}{E+m} (k_x \cos(kx)-k_y \sin(kx)) \\
+\frac{k_z}{E+m} \cos(kx) 
\end{array}\right) \quad .
\end{equation}
It is possible to simply define charge conjugation in the Majorana
representation by the complex conjugation of all spinor
components. In this case, the plane wave solutions
$m_{\pm \frac{1}{2}}(\vec{k},x)$ are obviously charge conjugation
eigenstates with eigenvalue $+1$. By construction, the eigenvalue
within the standard picture was chosen to be $-i$. This apparent
discrepancy is due to the fact that defining charge conjugation in
the Majorana representation by complex conjugation of the spinor
components implies a phase convention which is not in accordance
with the convention usually taken within the standard representation.
A complete analogy would be obtained by defining the charge conjugation
of a spinor $\psi_M$ in the Majorana representation via the standard
charge conjugation
\begin{equation}
\psi_S^c=i \gamma^2 \psi_S^*
\end{equation}
by
\begin{equation}
\psi_M^c=V \psi_S^c =V(i \gamma^2 (V^{-1} \psi_M)^*)=
i(V \gamma^2 V^{*-1}) \psi_M^*=-i \psi_M^*,
\end{equation}
which explains the differing eigenvalues obtained above.
Note that the charge conjugation matrix is not obtained by a basis
transform \`a la $\gamma_M^\mu=V \gamma^\mu V^{-1}$ due to the
antilinear nature of charge conjugation.

An alternative combination of particle and antiparticle solutions
would be
\begin{equation}
{\tilde{w}}_s(\vec{k},x)=\frac{1}{\sqrt{2}} (u_s(\vec{k},x)-i v_s(\vec{k},x)).
\end{equation}
The corresponding Majorana spinors ${\tilde{m}}_{\pm \frac{1}{2}}(\vec{k},x)$ are
then given by
\begin{equation}
{\tilde{m}}_{+\frac{1}{2}}(\vec{k},x)=
\left(\begin{array}{c}  \sin(kx)+\frac{1}{E+m} (k_x \cos(kx)+k_y \sin(kx)) \\
-\frac{k_z}{E+m} \cos(kx) \\
-\frac{k_z}{E+m} \sin(kx) \\
-\cos(kx) - \frac{1}{E+m} (k_x \sin(kx)-k_y \cos(kx))
\end{array}\right) \quad ,
\end{equation}
\begin{equation}
{\tilde{m}}_{-\frac{1}{2}}(\vec{k},x)=
\left(\begin{array}{c}  -\frac{k_z}{E+m} \cos(kx)  \\
+\sin(kx)-\frac{1}{E+m} (k_x \cos(kx)-k_y \sin(kx)) \\
+\cos(kx) - \frac{1}{E+m} (k_x \sin(kx)+k_y \cos(kx)) \\
+\frac{k_z}{E+m} \sin(kx) 
\end{array}\right) \quad .
\end{equation}
In the standard representation, the plane wave solutions
$w_s(\vec{k},x)$ and ${\tilde{w}}_s(\vec{k},x)$
are inequivalent.
At first sight, also ${\tilde{m}}_{\pm \frac{1}{2}}(\vec{k},x)$
seem to be of a different quality when compared to
$m_{\pm \frac{1}{2}}(\vec{k},x)$.
However, the ${\tilde{m}}_{\pm \frac{1}{2}}(\vec{k},x)$ are obtained
from the $m_{\pm \frac{1}{2}}(\vec{k},x)$ by the replacements
$\cos(kx) \rightarrow \sin(kx)$ and $\sin(kx) \rightarrow -\cos(kx)$,
or
\begin{equation}
\begin{array}{c}
\cos(kx) \rightarrow \cos(kx-\frac{\pi}{2}),\\
\sin(kx) \rightarrow \sin(kx-\frac{\pi}{2}),\\
\end{array}
\end{equation}
and therefore $m$ and ${\tilde{m}}$ are physically equivalent in the
sense that they can be related by a space-time translation,
which is not physically observable for states with sharply
defined momentum.

As a first conclusion, we observe that the real Majorana fields
discussed above transform according to a real representation of the
Lorentz group, whereas Dirac fields are related to both inequivalent
two-dimensional complex representations of the Lorentz group.
If we allow the spinors in the Majorana representation to become complex again,
all four spin-charge degrees of freedom of a Dirac spinor
which can be constructed as linear combinations from, e.g.,
$m_{\pm \frac{1}{2}}(\vec{k},x)$ and
$i {\tilde{m}}_{\pm \frac{1}{2}}(\vec{k},x)$
(the latter two states with negative C-parity) are introduced again, such that
the Dirac spinor describes charged particles again.

In is worthwhile to elaborate the Majorana case in further detail.
The well-known Dirac case requires no further discussion, since the four
independent solutions describing a particle with well-defined charge, spin and momentum
are readily constructed and given above, and a (plane wave) solution of the Dirac equation
\begin{equation}
i \gamma_\mu \partial^\mu \Psi_{_D}-m \Psi_{_D}=0 \label{Dirac_local}
\end{equation}
(like e.g. $u_s(\vec{k},x)$ or  $v_s(\vec{k},x)$)
multiplied by a complex phase remains a solution of the same equation.
This is not true in the Majorana case. One might be surprised why
we only found two physically distinct solutions of the Majorana equation from
the four plane wave solutions given above in the Dirac representation.

To resolve this situation, we first realize that multiplying
$m_{\pm \frac{1}{2}}(\vec{k},x)$ by the imaginary unit leads to charge eigenstates
with negative C-parity
\begin{equation}
[im_{\pm \frac{1}{2}}(\vec{k},x)]^*=[im_{\pm \frac{1}{2}}(\vec{k},x)]^c=
-im_{\pm \frac{1}{2}}(\vec{k},x).
\end{equation}
Yet, the plane wave solutions $im_{\pm \frac{1}{2}}(\vec{k},x)$ are no longer
solutions of the Majorana equation
\begin{equation}
i \gamma_\mu \partial^\mu \Psi_{_M}-m \Psi_{_M}^c=0 \label{Majo_local}
\end{equation}
but of
\begin{equation}
i \gamma_\mu \partial^\mu \Psi_{_M}+m \Psi_{_M}^c=0,
\end{equation}
since charge conjugation changes the sign of a purely imaginary spinor.
This defect can be easily resolved. Since $\gamma^5$ anticommutes with
the Dirac matrices $\gamma^5 \gamma^\mu=-\gamma^\mu \gamma^5$,
given an arbitrary solution $\Psi_{_D}$ of the Dirac equation eq. (\ref{Dirac_local}),
$\gamma^5 \Psi_{_D}$ is a solution of the Dirac equation with flipped mass
\begin{equation}
i \gamma_\mu \partial^\mu (\gamma^5 \Psi_{_D})+m (\gamma^5 \Psi_{_D})=0,
\end{equation}
and the same observation applies to the Majorana equation.
Modifying the spinors $w_s(\vec{k},x)$ given by eq. (\ref{Dirac_neutral})
according to
\begin{equation}
w_s(\vec{k},x) \rightarrow \hat{w}_s(\vec{k},x)=\gamma^5 w_s(\vec{k},x)
\end{equation}
and transforming $\hat{w}_s(\vec{k},x) \rightarrow V \hat{w}_s(\vec{k},x)$
into the Majorana representation yields two imaginary solutions
of the Majorana equation, since the multiplication of the original solutions
of the Dirac equation with $\gamma^5$ in conjunction with the fact that
the spinors in the Majorana representation are imaginary
leads to the correct sign of the mass term in the Majorana equation. An
explicit calculation yields
\begin{equation}
\hat{m}_s(\vec{k},x)= V \gamma^5 w_s(\vec{k},x),
\end{equation}
with
\begin{equation}
{\hat{m}}_{\frac{1}{2}}(\vec{k},x)= i
\left(\begin{array}{c}  -\frac{k_z}{E+m} \sin(kx)  \\
+\cos(kx)-\frac{1}{E+m} (k_x \sin(kx)-k_y \cos(kx)) \\
+\sin(kx) - \frac{1}{E+m} (k_x \cos(kx)+k_y \sin(kx)) \\
+\frac{k_z}{E+m} \cos(kx) 
\end{array}\right) \quad ,
\end{equation}
and
\begin{equation}
{\hat{m}}_{+\frac{1}{2}}(\vec{k},x)= i
\left(\begin{array}{c}  -\cos(kx)-\frac{1}{E+m} (k_x \sin(kx)+k_y \cos(kx)) \\
\frac{k_z}{E+m} \sin(kx) \\
\frac{k_z}{E+m} \cos(kx) \\
\sin(kx) + \frac{1}{E+m} (k_x \cos(kx)-k_y \sin(kx))
\end{array}\right) \quad .
\end{equation}

We finally conclude that the Majorana equation eq. ({\ref{Majo_local}) actually
describes four mass degenerate particle degrees of freedom with
definite four-momentum. These can be told apart,
e.g., by their charge conjugation eigenvalues, i.e. by requiring $\Psi_{_M}=\pm \Psi_{_M}$.
According to the transformation properties of the Majorana spinor under the
{\emph{real} spinor representation of the Lorentz group, solutions of the Majorana
equation can be combined linearly with {\emph{real} coefficients.
The Majorana equation itself does not force its solutions to be charge conjugation
eigenstates. It will become clear below, that it is even
possible that a four-component Majorana spinor describes Majorana particles
with two different masses.

The discussion presented so far is not the most general
one. In the case of the Dirac equation, charge conjugation (or rather CPT conjugation)
symmetry ensures that only one physical mass enters the equation. In the Majorana case,
a four-component spinor may represent a collection of two two-component Majorana
spinors with different masses. Furthermore, complex phases are possible in the most general
form of the Majorana equation, which then respects only the fundamental CPT symmetry.

\section{Majorana and Dirac mass terms} \label{masses}
\subsection{Lagrangian formulation}
In the foregoing sections, we introduced different fields
like the free (non-interacting) right- and left-chiral two-component Majorana spinors
$\chi_{_R}$ and $\chi_{_L}$, which fulfill in the most general case the massive wave equations
\begin{equation}
i (\sigma_0 \partial_0+\vec{\sigma} \vec{\nabla}) \chi_{_R}(x) -
\eta_{_R} m_{_R} \epsilon^{-1} \chi_{_R}^*(x)=0, \label{rightmass}
\end{equation}
\begin{equation}
i (\sigma_0 \partial_0-\vec{\sigma} \vec{\nabla}) \chi_{_L}(x) +
\eta_{_L} m_{_L} \epsilon^{-1} \chi_{_L}^*(x)=0, \label{leftmass}
\end{equation}
with real mass terms $m_{_{R,L}}$ and $|\eta_{_{R,L}}|=1$,
and, in the absence of complex mass phases, four-component Majorana spinors like
\begin{equation}
\nu_{_R}=\left(\begin{array}{c} \chi_{_R}  \\
0  \end{array}\right) \; , \quad
\nu_{_L}=\left(\begin{array}{c} 0  \\
\chi_{_L}  \end{array}\right) \; 
\end{equation}
or
\begin{equation}
\nu^{_M}_{_1}=
\left(\begin{array}{c}  \epsilon \chi^*_{_L} \\
\chi_{_L}  \end{array}\right)=
\left(\begin{array}{c}  \chi^*_{_{L_2}} \\ -\chi^*_{_{L_1}} \\
\chi_{_{L_1}} \\ \chi_{_{L_2}} \end{array}\right),
\qquad
\nu^{_M}_{_2}=
\left(\begin{array}{c}  \chi_{_R} \\ -\epsilon \chi^*_{_R} 
\end{array}\right)=
\left(\begin{array}{c}  \chi^{_1}_{_R} \\ \chi^{_2}_{_R} \\
-\chi^{_2*}_{_R} \\ \chi^{_1*}_{_R} \end{array}\right). \label{nu12}
\end{equation}
We have also seen that the fields above are basically equivalent,
i.e. from a physical point of view, it is irrelevant whether one uses
left- or right-chiral fields in a mathematical formalism
to describe a non-interacting Majorana field.
E.g., from the left-chiral field $\chi_{_L}$, which can be
projected out from $\nu^{_M}_{_1}$, one readily obtains
a right-chiral field $\epsilon \chi_{_L}^*$.
If $\chi_{_L}$ obeys eq. (\ref{leftmass}),
then $\epsilon \chi_{_L}^*$ obeys the equation
\begin{equation}
i (\sigma_0 \partial_0+\vec{\sigma} \vec{\nabla}) (\epsilon \chi_{_L}^*(x)) -
\eta_{_L}^* m_{_L} \epsilon^{-1} (\epsilon \chi_{_L}^*(x))^*=0, \label{rightmass}
\end{equation}
of course with the corresponding complex conjugate mass term, as one easily derives
by the help of the identity $\epsilon \vec{\sigma} \epsilon^{-1}=-\vec{\sigma}^*$.
Furthermore, how  $\nu^{_M}_{_1}$ and $\nu^{_M}_{_2}$ are connected also
becomes obvious by relating $\chi_{_{L_1}} \leftrightarrow -\chi^{_2*}_{_R}$ and
$\chi_{_{L_2}} \leftrightarrow  \chi^{_1*}_{_R}$.

We consider now two physically {\emph{different}} Majorana fields $\chi_{_R}$ and $\chi_{_L}$,
which are described by one left-chiral and a right-chiral two-component spinor field
for the sake of convenience.
The most general Lagrangian $\mathcal{L}$ which describes the free dynamics of these field
is given by $\mathcal{L}=\mathcal{L}_{_0}+\mathcal{L}_m$, where ($i \sigma_2=\epsilon$)
\begin{equation}
\mathcal{L}_{_0}  =  \chi_{_L}^+ i \bar{\sigma}^\mu \partial_\mu \chi_{_L}
+\chi_{_R}^+ i \sigma^\mu \partial_\mu \chi_{_R},
\end{equation}
and 
\begin{eqnarray}
\mathcal{L}_m & = &
-\eta_{_L} \frac{m_{_L}}{2} \chi_{_L}^+ i \sigma_2 \chi_{_L}^* 
+\eta_{_L}^* \frac{m_{_L}}{2} \chi_{_L}^{\mbox{\tiny{T}}} i \sigma_2 \chi_{_L} \nonumber \\
& & +\eta_{_R} \frac{m_{_R}}{2} \chi_{_R}^+ i \sigma_2 \chi_{_R}^* 
-\eta_{_R}^* \frac{m_{_R}}{2} \chi_{_R}^{\mbox{\tiny{T}}} i \sigma_2 \chi_{_R} \nonumber\\
& & - \eta_{_{D}} m_{_{D}} \chi_{_L}^+ \chi_{_R} 
- \eta_{_{D}}^* m_{_{D}} \chi_{_R}^+ \chi_{_L},
\end{eqnarray}
where the $m_{D(irac)}$-terms couple the left- and right-chiral fields, as it is the case
in the Dirac equation.
Note that all the terms above transform as scalars. E.g., having a glimpse at eqns. (\ref{trafo_law})
and (\ref{trafo_law_right}) and neglecting space-time arguments for the moment, one finds
\begin{equation}
\chi_{_R} \rightarrow S \chi_{_R},
\quad \chi_{_L} \rightarrow \epsilon S^* \epsilon^{-1} \chi_{_L},
\end{equation}
and thus indeed
\begin{equation}
\chi_{_L}^+ \chi_{_R} \rightarrow \chi_{_L}^+ \epsilon S^{\mbox{\tiny{T}}}
\epsilon^{-1} S \chi_{_R}=\chi_{_L}^+ \epsilon \epsilon^{-1} S^{-1} \epsilon
\epsilon^{-1} S \chi_{_R}=\chi_{_L}^+ \chi_{_R}.
\end{equation}

For the following discussion, we require first that $m_{_L} \neq 0$ or $m_{_R} \neq 0$.
Before coming back to the Lagrangian itself, we analyze the wave equation which follows from
the Lagrangian. One has
\begin{equation}
i \left(\begin{array}{cc}
0 & \sigma_0 \partial_0 -\vec{\sigma} \vec{\nabla} \\
\sigma_0 \partial_0 +\vec{\sigma} \vec{\nabla} & 0  
\end{array}\right)
\left(\begin{array}{cc}
\chi{_R} \\ \chi{_L} 
\end{array}\right)-
\left(\begin{array}{cc}
\eta_{_D} m_{_D} & +\eta_{_L} m_{_L} \epsilon K \\ -\eta_{_R} m_{_R}  \epsilon K & \eta_{_D}^*
m_{_D}  \end{array}\right)
\left(\begin{array}{cc}
\chi_{_R} \\ \chi{_L} 
\end{array}\right)=0, \label{general1}
\end{equation}
where $K$ denotes complex conjugation.
Introducing the four-spinor 
\begin{equation}
\Psi=
\left(\begin{array}{c}
\chi_{_R} \\ \chi_{_L}
\end{array}\right) \quad ,
\end{equation}
eq. (\ref{general1}) can now be written
\begin{equation}
i \gamma^\mu \partial_\mu \Psi-\tilde{m}_{_M} \Psi^c - \tilde{m}_{_D} \Psi=0 \label{general2}
\end{equation}
with appropriately chosen mass matrices $\tilde{m}_{_M}$ and $\tilde{m}_{_D}$
\begin{equation}
\tilde{m}_{_M}=\left(\begin{array}{cc}
\eta_{_L} m_{_L} & 0 \\ 0  &  \eta_{_R} m_{_R}
\end{array}\right), \quad
\tilde{m}_{_D}=\left(\begin{array}{cc}
\eta_{_D} m_{_D} &  0 \\ 0 & \eta_{_D}^* m_{_D}  \end{array}\right).
\end{equation}
since charge conjugation is given in the chiral representation by
\begin{equation}
C[\Psi]=\Psi^c=\tilde{\epsilon} \Psi^*=i \gamma^2 \Psi^*=i \gamma^2 \gamma^0
\bar{\Psi}^{\mbox{\tiny{T}}},
\quad \tilde{\epsilon}=
\left(\begin{array}{cc}
0 & \epsilon \\ -\epsilon & 0
\end{array}\right).
\end{equation}

The present two-neutrino theory contains three phases $\eta_{_D}$ and $\eta_{_{L,R}}$.
Though, two phases can be eliminated (i.e., set equal to one)
by multiplying $\chi_{_{L,R}}$ with appropriate phases, respectively, without
changing the physical content of the theory. E.g.,
as we have seen above, it is possible to remove the Dirac phase $\eta_{_D}$
by a chiral transformation of the spinor according to eq. (\ref{chiral_gauge}).
A subsequent multiplication of both $\chi_{_{L,R}}$ by the same phase can then be
used modify $\eta_{_{L}}$ or $\eta_{_{R}}$, but in general, a CP violating phase
will survive. Vice versa, it is possible to remove the phases $\eta_{_L}$ and $\eta_{_R}$,
then possibly a non-trivial phase $\eta'_{_D}$ survives.
How the respective phases are intertwined can be most easily inferred from
the invariance of the mass Lagrangian $\mathcal{L}_m$ under a phase redefinition
of the fields. 
In the special case where $m_{_D}=0$, all phases can be removed and the wave equation describes to
Majorana particles, and the case where $\tilde{m}_{_M}=0$ leads to the usual Dirac theory,
as follows directly from eq. (\ref{general2}).
We now examine two different scenarios in further detail. 

\subsection{CP symmetric theory with Majorana masses}
In the case $\tilde{m}_{_M} \neq 0$,
we may introduce new four-component fields
inspired by eq. (\ref{nu12})
\begin{equation}
\nu_{_1}=
\left(\begin{array}{c}  \eta_{_L} \epsilon \chi^*_{_L} \\
\chi_{_L}  \end{array}\right),
\qquad
\nu_{_2}=
\left(\begin{array}{c} \chi_{_R} \\
- \eta_{_R}  \epsilon \chi^*_{_R} 
\end{array}\right),
\end{equation}
such that we obtain in the chiral representation ($\epsilon^2=-1$)
\begin{equation}
\nu_{_1}^c=
\left(\begin{array}{cc}  0 & \epsilon \\
-\epsilon & 0  \end{array}\right)
\left(\begin{array}{c}  \eta_{_L} \epsilon \chi^*_{_L} \\
\chi_{_L}  \end{array}\right)^*=
\left(\begin{array}{c}  \epsilon \chi^*_{_L} \\
\eta_{_L}^* \chi_{_L}  \end{array}\right)=
\eta_{_L}^*
\left(\begin{array}{c} \eta_{_L} \epsilon \chi^*_{_L} \\
\chi_{_L}  \end{array}\right),
\end{equation}
\begin{equation}
\nu_{_2}^c=
\left(\begin{array}{cc}  0 & \epsilon \\
-\epsilon & 0  \end{array}\right)
\left(\begin{array}{c} \chi_{_R} \\
- \eta_{_R} \epsilon  \chi^*_{_R} 
\end{array}\right)^*=
\left(\begin{array}{c} \eta_{_R}^* \chi_{_R} \\
-\epsilon  \chi^*_{_R} \end{array}\right) =
\eta_{_R}^*
\left(\begin{array}{c} \chi_{_R} \\
-\eta_{_R} \epsilon \chi_{_R}^*  \end{array}\right),
\end{equation}
i.e. $\nu_{_1}$ and  $\nu_{_2}$ are charge conjugation eigenstates.
In addition, we introduce the Dirac adjoint spinors ($\epsilon^{\mbox{\tiny{T}}}=-\epsilon$)
\begin{equation}
\bar{\nu}_{_1}={\nu}_{_1}^+ \gamma^0_{ch}=
\left(\begin{array}{cc}  \eta_{_L}^* \chi^{\mbox{\tiny{T}}}_{_L}
\epsilon^{\mbox{\tiny{T}}} & \chi_{_L}^+ \end{array}\right)
\left(\begin{array}{cc}  0 & 1 \\ 1 & 0  \end{array}\right)=
\left(\begin{array}{cc} \chi_{_L}^+ &
- \eta_{_L}^* \chi^{\mbox{\tiny{T}}}_{_L} \epsilon
\end{array}\right)
\end{equation}
and
\begin{equation}
\bar{\nu}_{_2}={\nu}_{_2}^+ \gamma^0_{ch}=
\left(\begin{array}{cc}  \chi^+_{_R} &
-\eta_{_R}^* \chi^{\mbox{\tiny{T}}}_{_R} \epsilon^{\mbox{\tiny{T}}} \end{array}\right)
\left(\begin{array}{cc}  0 & 1 \\ 1 & 0  \end{array}\right)=
\left(\begin{array}{cc}  \eta_{_R}^*  \chi_{_R}^{\mbox{\tiny{T}}} \epsilon &
\chi^+_{_R}
\end{array}\right).
\end{equation}
Next, we observe that
\begin{equation}
\bar{\nu}_{_1} {\nu}_{_1}=
+\eta_{_L} \chi_{_L}^+ \epsilon \chi_{_L}^*
-\eta_{_L}^* \chi_{_L}^{\mbox{\tiny{T}}} \epsilon \chi_{_L}, \quad
\end{equation}
\begin{equation}
\bar{\nu}_{_2} {\nu}_{_2}=
-\eta_{_R} \chi_{_R}^+ \epsilon \chi_{_R}^*
+\eta_{_R}^* \chi_{_R}^{\mbox{\tiny{T}}} \epsilon \chi_{_R}, \quad
\end{equation}
and
\begin{equation}
\bar{\nu}_{_1} {\nu}_{_2}=
\chi_{_L}^+ \chi_{_R}
-\eta_{_L}^* \eta_{_R} \chi_{_L}^{\mbox{\tiny{T}}} \chi_{_R}^*=
\chi_{_L}^+ \chi_{_R}+ \eta_{_L}^* \eta_{_R} \chi_{_R}^+ \chi_{_L},
\end{equation}
\begin{equation}
\bar{\nu}_{_2} {\nu}_{_1}=(\bar{\nu}_{_1} {\nu}_{_2})^+=
\chi_{_R}^+ \chi_{_L}
+ \eta_{_R}^* \eta_{_L} \chi_{_L}^+ \chi_{_R},
\end{equation}
where we used the fact that the fields are anticommuting (Grassmann) numbers.

As discussed above, we are free to gauge $\chi_{_{L,R}}$ with appropriate phases
so that $\eta_{_L}=\eta_{_R}=1$. Generally, a (modified) Dirac phase $\eta'_{_D}$
will  survive. {\emph{Only}} when this phase is trivial, i.e., for $\eta'_{_D}=1$
the mass term $\mathcal{L}_m$ can be written by the help of self-charge conjugate
fields $\bar{\nu}_{_{1,2}}$
\begin{equation}
\mathcal{L}_m= -\frac{1}{2}
\left(\begin{array}{cc}  \bar{\nu}_{_1} & \bar{\nu}_{_2} \end{array}\right)
\left(\begin{array}{cc}  m_{_L} & m_{_D} \\ 
m_{_D} & m_{_R}  \end{array}\right)
\left(\begin{array}{c}  {\nu}_{_1} \\ {\nu}_{_2} \end{array}\right).
\end{equation}
where now $C[\nu_{_{1,2}}]=\nu_{_{1,2}}^c=+\nu_{_{1,2}}$.
To be more specific, the trick above works if $\eta_{_L}^* \eta_{_R} \eta_{_D}^2=1$.

Since the mass matrix above is real and symmetric, a real linear combination
$\nu'_{_{1,2}}$ of the fields $\nu_{_{1,2}}$ can be found, which is again
self-charge conjugate, so that the mass matrix becomes diagonal.
Of course, the mass of the neutral fields $\nu'_{_{1,2}}$ is given by
the eigenvalues $m_{_{1,2}}$ of the mass matrix (note the normalization factor). From
\begin{equation}
(\lambda-m_{_1})(\lambda-m_{_2})=(\lambda-m_{_L})(\lambda-m_{_R})-m_{_D}^2,
\end{equation}
we obtain
\begin{equation}
m_{_{1,2}}=\frac{m_{_L}+m_{_R}}{2} \pm \sqrt{ \biggl( \frac{m_{_L}-m_{_R}}{2} \biggr)^2
+m_{_D}^2 }.
\end{equation}
We conclude that in the presence of Dirac {\emph{and}} Majorana mass terms and in
the absence of CP violating phases, the Lagrangian $\mathcal{L}_m$ describes
two charge neutral Majorana particles.
This special case corresponds to the discussion originally given in \cite{Cheng}, which is now
widely found in the literature. For a discussion in the Lagrangian framework of the
Dirac field as the degenerate limit of $m_{_L}=m_{_R}=0$ in the more general case
of two Majorana particles, we also refer to \cite{Cheng}.
 
\subsection{CP violating theory with Majorana masses}
The case where a phase remains in the theory is more involved.
In order to motivate the discussion below, we point out that
that in the general Lagrangian $\mathcal{L}$ defined above,
two Majorana fields which transform according to the left- and
right-chiral representation of the Lorentz group were used.
However, in extensions of the Standard Model, it is usually assumed
that the Higgs mechanism generates mass terms for leptons like
the electron and muon neutrino, which can be written within a simplified
model in several equivalent forms, e.g.
\begin{equation}
-\mathcal{L}_m^{e \mu} \sim 
\left(\begin{array}{cc}  {\nu}_{_e}^{\mbox{\tiny{T}}} & {\nu}_{_\mu}^{\mbox{\tiny{T}}}
\end{array}\right)_{_L} \tilde{C} \tilde{M}_{_{e \mu}}
\left(\begin{array}{c}  {\nu}_{_e} \\ {\nu}_{_\mu} \end{array}\right)_{_L} + h.c.,
\end{equation}
where $\tilde{C}=i \gamma^2 \gamma^0$ is a charge conjugation matrix
($-\tilde{C}=\tilde{C}^{\mbox{\tiny{T}}} = \tilde{C}^+ = \tilde{C}^{-1}$)
and $\tilde{M}_{_{e \mu}}$ is a complex and symmetric
$2 \times 2$ matrix in flavor space
\begin{equation}
\tilde{M}=
\left(\begin{array}{cc}  m_{ee} e^{i \gamma_e} & m_{e \mu} e^{i \gamma_{e \mu}} \\
m_{e \mu} e^{i \gamma_{e \mu}} & m_{\mu \mu} e^{i \gamma_{\mu}} 
\end{array}\right).
\end{equation}
So it is common practice that both the electron neutrino and the muon neutrino
are described by left-chiral fields. Since in the present formalism,
$\epsilon \chi_{_R}^*$ transforms accordingly, we describe the mass
terms in the Lagrangian $\mathcal{L}_m$ in a similar manner as above
\begin{equation}
\mathcal{L}=-\frac{1}{2}
\left(\begin{array}{cc}  \chi_{_L}^+  & \chi_{_R}^{\mbox{\tiny{T}}} \epsilon
\end{array}\right) M
\left(\begin{array}{c}  \epsilon \chi_{_L}^* \\ \chi_{_R} \end{array}\right)
=
-\frac{1}{2}
\left(\begin{array}{cc}  \chi_{_L}^+  & \chi_{_R}^{\mbox{\tiny{T}}} \epsilon
\end{array}\right)
\left(\begin{array}{cc}  m_{_L}  & \eta_{_D} m_{_D} \\
\eta_{_D} m_{_D} &  m_{_R}
\end{array}\right)
\left(\begin{array}{c}  \epsilon \chi_{_L}^* \\ \chi_{_R} \end{array}\right)
+ h.c., \label{massmatrix}
\end{equation}
where we made use of the fact that that it is possible to eliminate
two of three phases in a two-Majorana particle theory. Note that in eq. (\ref{massmatrix}),
one has $\chi_{_R}^{\mbox{\tiny{T}}} \chi_{_L}^*=
-(\chi_{_R}^{\mbox{\tiny{T}}} \chi_{_L}^*)^{\mbox{\tiny{T}}}=
-\chi_{_L}^+ \chi_{_L}$ due to the Fermi statistics, therefore the mass
matrix is symmetric, but not Hermitian.

A short calculation shows that the unitary matrix
\begin{equation}
U=\left(\begin{array}{cc}  \cos \vartheta & -\sin \vartheta e^{ i \alpha} \\
\sin \vartheta e^{-i \alpha} &  \cos \vartheta
\end{array}\right)
\end{equation}
diagonalizes $M$
\begin{equation}
U^{\mbox{\tiny{T}}} M U = \left(\begin{array}{cc}  \hat{m}_1  & 0 \\
0 &  \hat{m}_2
\end{array}\right)
\end{equation}
if $\vartheta$ and $\alpha$ fulfill the constraint
\begin{equation}
2 \eta_{_D} m_{_D} \cos (2 \vartheta)= \sin(2 \vartheta) ( m_{_L} e^{i \alpha} -
m_{_R} e^{-i \alpha}).
\end{equation} 
The Majorana fields
\begin{equation}
\left(\begin{array}{c}  \epsilon \chi_{_L}^{'*} \\ \chi'_{_R} \end{array}\right)
= U^+ \left(\begin{array}{c}  \epsilon \chi_{_L}^* \\ \chi_{_R} \end{array}\right)
\end{equation}
would diagonalize the mass Lagrangian, but since $\hat{m}_1$ and $\hat{m}_2$ are
complex, they do not yet correspond to the "physical states" in our model.
However, writing $\hat{m}_{1,2}=e^{i \Phi_{1,2}}$, we have
\begin{equation}
U^{\mbox{\tiny{T}}} M U = \Phi \left(\begin{array}{cc}  \hat{m}_1  & 0 \\
0 &  \hat{m}_2\end{array}\right)  \Phi ,
\end{equation}
where
\begin{equation}
\Phi=\left(\begin{array}{cc}  e^{i \Phi_1/2}  & 0 \\
0 &  e^{i \Phi_2/2}   \end{array}\right).
\end{equation}
$U'=U \Phi$ brings the mass matrix into diagonal form with real and positive
eigenvalues. These can be interpreted as the physical mass of the neutrino
fields.

The present discussion is the starting point of many models which aim
at a description of lepton mixing and mass hierarchies, like, e.g., the
seesaw mechanism \cite{seesaw1,seesaw2}. In realistic models with three-generation
Majorana neutrinos, even two non-trivial Majorana phases appear, which are related
to the important issue of CP nonconservation \cite{Bilenky}.

We finally comment (once more) on the Majorana fields constructed in this section.
It is a simple exercise to demonstrate that
the CPT transformed four-component spinor $\Psi$ in eq. (\ref{general2})
\begin{equation}
CPT[\Psi(x)]=\Psi^{cpt}=i \gamma^5 \Psi(-x)= i \gamma^5 \Psi(-x^0,-\vec{x})
\end{equation}
again fulfills eq. (\ref{general2}), but the CP symmetry is broken
when Majorana masses and a CP violating phase are inherent in the theory.
CPT transforms spin and momentum of a particle. Whereas the spin changes
sign, the momentum is flipped twice by the P and the T operation and remains
invariant. However, a massive CPT transformed Majorana particle state can be transformed back
into the original state by an appropriate Poincar\'e transformation.
The helicity of the particle is not preserved in this process and changes sign,
but the chirality, i.e. the transformation property of the field which describes the
particle remains invariant. The situation is completely
different for the degenerate case of Dirac particles, where CPT changes the
charge of the particle, which is a Lorentz invariant quantity.

\end{document}